\documentclass[aps,prx,superscriptaddress,floatfix,amsmath,amsfonts,twocolumn]{revtex4-2}  
\usepackage[russian,english]{babel}

\usepackage{enumitem}
\usepackage{amsmath}
\usepackage{graphicx}
\usepackage{epsfig}
\usepackage{bm}
\usepackage{bbold}
\usepackage{amssymb}
\usepackage{color}
\usepackage[normalem]{ulem}  %

\newcommand{\cM}{{\cal M}}

\begin{document}

\title{Vorticity of polariton condensates in rotating traps. } 

\author{A.V. Yulin}
\affiliation{Department of Physics, ITMO University, Saint Petersburg 197101, Russia}
\affiliation{Science Institute, University of Iceland, Dunhagi 3, IS-107, Reykjavik, Iceland}

\author{I.A. Shelykh}
\affiliation{Science Institute, University of Iceland, Dunhagi 3, IS-107, Reykjavik, Iceland}
\affiliation{Department of Physics, ITMO University, Saint Petersburg 197101, Russia}
\affiliation{Abrikosov Center for Theoretical Physics, MIPT, Dolgoprudnyi, Moscow Region 141701, Russia}

\author{E. S. Sedov}
\affiliation{Russian Quantum Center, Skolkovo, Moscow 143025, Russia}
\affiliation{Spin-Optics laboratory, St. Petersburg State University, St. Petersburg 198504, Russia}
\affiliation{Vladimir State University, Vladimir 600000, Russia}

\author{A.V. Kavokin}
\affiliation{Westlake University, School of Science, 18 Shilongshan Road, Hangzhou 310024, Zhejiang Province, China}
\affiliation{Westlake Institute for Advanced Study, Institute of Natural Sciences, 18 Shilongshan Road, Hangzhou 310024, Zhejiang Province, China}
\affiliation{Abrikosov Center for Theoretical Physics, MIPT, Dolgoprudnyi, Moscow Region 141701, Russia}

\date{\today}

\begin{abstract}

This work is inspired by recent experiments on the formation of vortices in exciton-polariton condensates placed in rotating optical traps. We study theoretically the dynamics of formation of such vortices and elucidate the fundamental role of the mode competition effect in determining the properties of stationary polariton states triggered by stimulated scattering of exciton-polaritons. The interplay between linear and non-linear effects is shown to result in a peculiar polariton dynamics. However, near the lasing threshold, the predominant contribution of the nonlinear effects is the saturation of the linear gain.  

 \end{abstract}

\maketitle

\section{Introduction}

Semiconductor systems suitable for the realization of strong light-matter coupling have been actively studied in the recent years \cite{Carusotto2013}. The reason for the interest they attract is the hybridization between the cavity photons and electronic excitations, which gives rise to the appearance of quasiparticles having extremely low effective masses and able to efficiently interact with each other. Probably the most remarkable achievement in this field is the experimental realization of Bose-Einstein condensation of exciton polaritons at extraordinarily high temperatures \cite{kasprzak2006bose,balili2007bose}. This fundamental discovery paved the way for practical realization of polariton lasers  \cite{pau1996observation,christopoulos2007room,schneider2013electrically}.

Coherent ensembles of interacting polaritons reveal the phenomenon of superfluidity \cite{Carusotto2013}, and in certain geometries polariton states sustaining persistent currents may be formed \cite{kavokin2003polariton,carusotto2004probing,cristofolini2013optical,chestnov2016nonlinear,lerario2017room,Lukoshkin2018,Sedov2021,SciRep1122382,Lukoshkin2023,sanvitto2010persistent}. The corresponding nonlinear localized structures characterized by a topological charge, known as polariton vortices, have been extensively studied from both theoretical and experimental perspectives~\cite{lagoudakis2008quantized,lagoudakis2009observation,sanvitto2010persistent,nardin2011hydrodynamic,lagoudakis2011probing}. 

States of the opposite vorticity can form a basis of a polariton qubit \cite{Xue2021,Kavokin2022}. It is therefore important to possess a tool to have control over  the direction of the polariton rotation. This can be achieved by using chiral structures~\cite{assmann2012all,dall2014creation}, fine tuning of the excitation conditions \cite{nardin2010selective}, use of the effect of spin to angular momentum conversion \cite{manni2011spin}, use of Laguerre-Gaussian beams \cite{kwon2019direct} or application of external magnetic fields~\cite{yulin2020spinning}.

The study of mesoscopic coherent polariton states, including vortices, confined in the microcavity plane using potential traps 
has attracted considerable interest~\cite{PhysRevB74155311,1010634983832,NatPhys6860,LSA879,PhysRevB91195308,PhysRevB97235303,PhysRevB101245309,PhysRevB107045302}. Particularly noteworthy are annular optical traps, which can be induced in the microcavity by laser beams with appropriate spatial profiles~\cite{PhysRevB91195308,PhysRevB97235303,PhysRevB101245309,PhysRevB107045302}.
It was recently shown that formation of an external rotating potential is a powerful tool of control of the polariton states forming in annular traps~\cite{lagoudakis2022,Fraser}. 
The goal of this paper is to provide the theoretical analysis of the dynamics of the polariton condensates in these geometries in the vicinity of a polariton lasing threshold and to elucidate the mechanism responsible for the symmetry breaking between the clock and counter-clockwise rotating solutions. We consider two distinct experimentally relevant situations, corresponding to three different coupling schemes between the states of distinct vorticities induced by the rotating potential.
 Vorticity is characterized by the angular index $m$ that quantifies the number of times the phase of a wave winds as one moves around the vortex core.
Namely, we analyse the cases of the coupling between $m=0$ and $m=\pm 1$ states, the coupling between $m=\pm 1$ and $m=\pm 2$ states and the coupling between $m=-1$ and $m=+1$ states \cite{lagoudakis2022}. 

The formation of polariton vortices in a trap with a rotating complex potential is considered within the framework of the generalized scalar Gross-Pitaevskii equation. It is shown that close to the polariton lasing threshold, the stationary states inherit their angular momentum from the fastest growing linear mode. It is worth noting that in the presence of the azimuth non-symmetric potential, the polariton eigenmodes can be expressed as a combination of multiple components with different angular indices~\cite{Sedov2021,SciRep1122382}. The resulting direction of rotation of the polariton state is determined by the relative weights associated with these components. These weights, in turn, are influenced by the angular velocity of the rotating potential. At certain threshold angular velocities, the velocity of the condensate can cross zero and undergo a change in sign.

\begin{figure}[tb!]
\begin{center}
\includegraphics[width=.95\linewidth]{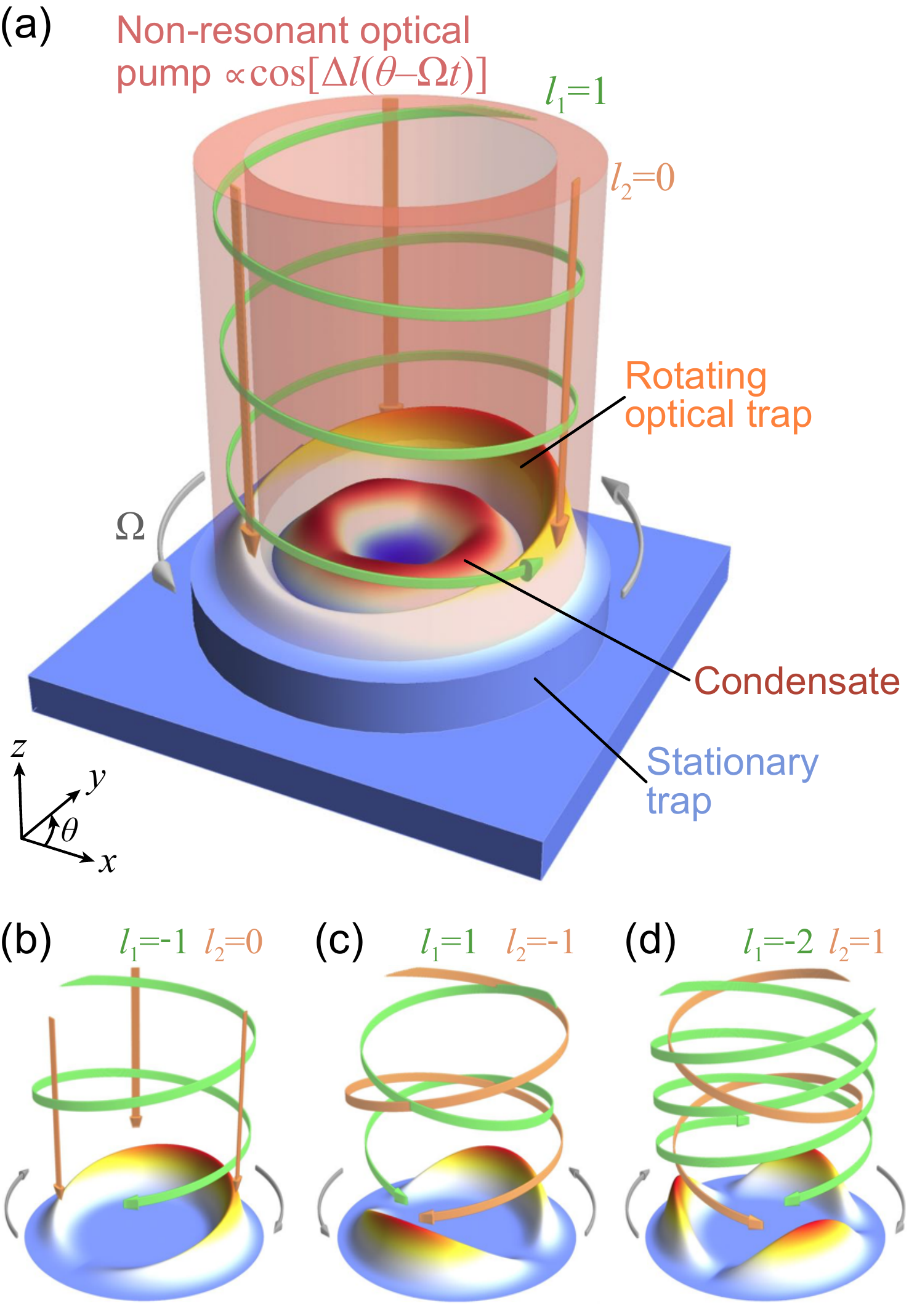}
\end{center}
\caption{ \label{FIG_Scheme}
(a) Schematic depicting the excitation of a polariton condensate in an external rotating optical potential.
The rotation is induced by two laser beams with angular momenta  $l _1 = 1$ and $l _2 = 0$, characterized by different frequencies~$\Omega=\omega_1-\omega_2$.
(b--d)~The shape of the rotating part of the pump-induced optical trap depending on the angular momenta of the optical pump components.
The frequencies in the panels are taken as $\omega _1 > \omega _2$.
}
\end{figure}

There is an additional scenario that can lead to a change in the rotation velocity of the condensate. 
This scenario arises from a swap between the fastest and the second fastest growing modes. In the case considered in the paper, the polaritons in these modes rotate in different directions, with not only different absolute values of velocities but also opposite signs. Therefore, if the fastest and the second fastest modes swap  at certain velosity, the rotation direction of the stationary polaritons also changes. In this scenario, the velocity of the polariton rotation is not a continuous function of the rotating potential velocity. Instead, it undergoes a distinct change by a finite value at critical velocities of the potential.  
To analytically describe the observed effects, a perturbation approach was employed. The developed coupled modes theory effectively reproduces the results obtained from direct two-dimensional (2D) simulations.

The paper is organized as follows. After the introduction, in section II we describe the considered geometry and introduce the corresponding mathematical model. In the third section, we provide a brief discussion of the formation of the hybrid states in the presence of the rotating potential that couples modes with different angular indices. Section IV contains the numerical results concerning the hybridization of the modes with an angular index difference equal to~$1$. The perturbation theory explaining the numerical results is developed in the fifth section. Various mechanisms governing the formation of stationary polariton states in the rotating potentials are discussed in Section VII. Finally, in the Conclusion we briefly summarize the main finding of the paper.

\section{The system in question and its mathematical description}

We consider the geometry of an annular trap created by an axially symmetric incoherent pump which produces both the conservative confining potential for a condensate due to the effect of reservoir-induced blueshift, and the effective linear gain provided by the stimulated condensation of the polaritons from a reservoir to a coherent state. We also account for an axially symmetric conservative potential created by microstructuring of a sample. 
A schematic representation of the excitation and trapping of the polariton condensate is shown in Fig.~\ref{FIG_Scheme}(a).

Control of vorticity is achieved via additional rotating non-resonant pump, which can be created by application of the two interfering optical beams beams with different orbital angular momenta (Laguerre-Gaussian modes) and slightly detuned frequencies and thus reads:
\begin{equation}
P_r (\mathbf{r},t)=\tilde f(r)\cos[\Delta l (\theta-\Omega t)],\label{Prt}   
\end{equation}
where $\tilde f(r)$ describes the radial dependence of the pump, $\Delta l = l_1 - l_2$ denotes the difference in the angular indices between the beams creating the pump, $\Omega = \omega _1 - \omega _2$ is the rotation velocity of the potential defined by the detuning between the two pumping modes, $r$ and $\theta$ correspond to the radial and angular coordinates, respectively.
Indices 1 and 2 enumerate the interfering beams.
In panels (b--d) of Fig.~\ref{FIG_Scheme}, we show schematically examples of the rotating potentials created by two beams with different combinations of $l_1$ and $l_2$. We assume $\omega_1 > \omega_2$ for definiteness. 
The absolute value of the difference $\Delta l$ determines the number of variations of the potential in the azimuthal direction, while its sign determines the direction of rotation of the potential.
It is worth noting that when $\omega _1 < \omega _2$, the rotation direction of the potential trap is reversed.

Let us stress once more, that the rotating pump results both the rotating conservative potential and rotating gain for the condensate, as it will become clear from the dynamic equations that we introduce below.

In the scalar case, the problem can be described by the equations, that govern the polariton condensate characterized by the order parameter function~$\Psi(\textbf{r},t)$ and the density of the incoherent exciton reservoir~$\rho(\textbf{r},t)$:
\begin{widetext}
\begin{subequations}
\label{master_eqs}
\begin{eqnarray}
&& i \hbar \partial_{t} \Psi = \frac{\hbar^2}{2m^*} \nabla^2 \Psi-\left[ \tilde V(r) + i\hbar \Gamma_p + \hbar (g_2-i g_1) \rho \right] \Psi- (H+i \tilde H)|\Psi|^2\Psi, \label{master_eqPol} \\
&&\partial_{t} \rho=-(\Gamma_r+2 g_1 |\Psi|^2) \rho +P(\mathbf{r}, t), \label{master_eqRes}
\end{eqnarray}
\end{subequations}
\end{widetext}
where $m^*$ is the effective mass of polaritons in the microcavity plane,  $\tilde V(r)$ is the axially symmetric conservative potential created by microstructuring, the term $\hbar \Gamma_p \psi$ accounts for the intrinsic losses in the cavity that we assume to be spatially uniform. The coefficient $g_1$ is the rate of the stimulated scattering of reservoir excitons into the coherent polariton state, the coefficient $g_2$ defines the reservoir induced blueshift of the condensate as the function of its density. The nonlinear self-induced blueshift of the coherent polariton state is described by the parameter $H$, and the coefficient $\tilde H$ accounts for the nonlinear losses of coherent polaritons. Finally, $\Gamma_r$ is the relaxation rate of the  incoherent reservoir excitons. The total incoherent pump is accounted for by $P(\mathbf{r},t)$ standing in the right hand side of~Eq.~\eqref{master_eqRes}. The pump $P(\mathbf{r},t)=P_0(r) + P_r(\mathbf{r},t)$ contains two components, including the time independent axially symmetric term $P_0(r)$ and the rotating term $P_r (\mathbf{r},t)$ defined by Eq.~\eqref{Prt}.  

Let us assume that the relaxation of incoherent excitons is fast compared to other characteristic timescales of the polariton dynamics. This allows us to adiabatically eliminate $\rho$ putting $\partial\rho/\partial t=0$ and thus  get for the following equation for the macroscopic wavefunction of a condensate. Introducing the dimensionless time $t \rightarrow \left. t \right/ 2\Gamma_p$, coordinate $x \rightarrow x \sqrt{\hbar / 2m ^*\Gamma_p}$ and the order parameter function $\psi = \Psi\sqrt{2g_1 / \Gamma_r}$, we get
\begin{widetext}
\begin{eqnarray}
i\partial_t \psi =  \frac{1}{2} \nabla^2 \psi  - \left(V +\frac{i}{2} \right)\psi  - (1-i\alpha)  \frac{W_0+W_r}{1 +  |\psi|^2} \psi - (h+i \tilde h)|\psi|^2\psi, \label{master_scalar}
\end{eqnarray}
\end{widetext}
where $V={\tilde V}/{2\hbar \Gamma_p}$, $W_0={P_0 g_2}/{2\Gamma_p \Gamma_r}$ and $W_r= {P_r g_2} / {2 \Gamma_p \Gamma_r}= f(r)\cos[\Delta l (\theta-\Omega t)]$ are the effective stationary and rotating potentials created by the pump with $f={\tilde f g_2}/{2 \Gamma_p \Gamma_r}$,  and $\Omega$ is given in units of $2 \Gamma _p$.

$\alpha={g_1}/{ g_2}$ is the ratio of the gain and pump-induced blueshift,  $ h={H \Gamma_r}/{4 \hbar g_1 \Gamma_p}$ and $\tilde h={\tilde H \Gamma _r}/{4 \hbar g_1 \Gamma_p}$ are normalized coefficients that account for the real and imaginary parts of the  polariton nonlinearity.

\section{Hybridization of the states in rotating potentials}

In this section, we consider the scenario where the rotating potential couples the scalar states with angular indices $m=0$ and $m =\pm 1$. This can be realizes, if external non-resonant pump is created by the superposition of a simple Gaussian beam and a Laguerre-Gaussian beam with $l=\pm 1$. 
The shape of the rotating part of such potential is schematically shown in Fig.~\ref{FIG_Scheme}(b).

The linearized equation (\ref{master_scalar}) that describes the dynamics of the condensate close to the lasing threshold reads:
\begin{widetext}
\begin{eqnarray}
\label{EqGPELinearized}
    i\partial_t \psi - \frac{1}{2} \nabla^2 \psi  +\left( \frac{i}{2}+V \right)\psi  + (1-i\alpha)  W_0 \psi= (i\alpha- 1)W_r \psi.  \label{master_scalar_lnr}
\end{eqnarray}
\end{widetext}
Let us assume the rotating potential to be a small correction. This is physically relevant case that allows us to develop a simple perturbation theory giving a physical insight into the polariton dynamics.

The solution of the unperturbed equation~\eqref{EqGPELinearized} with the right hand side set to zero can be expanded as
\begin{equation}
\label{app_anzats}
\psi=\sum_{m, q} C_{m, q} (t) \psi_{m, q},
\end{equation}
where $\psi_{m, q}= R_{m, q}(r) \exp{im\theta}$ are eigenfunctions of the azimuthally symmetric problem corresponding to the eigenvalues $\lambda_{m, q}$ with functions $R_{m, q}$ describing the radial condensate distribution, $m$ and $q$ are angular and radial quantum numbers. For convenience, we use the normalized eigenfunctions, $\int |\psi_{m, q}|^2 dx dy=1$. The amplitude coefficients $C_{m, q}$ are governed by the equations 
\begin{eqnarray}
\label{EqLinearizedUnpertEq}
    \dot C_{m, q}=\lambda_{m, q} C_{m, q}. \label{q}
\end{eqnarray}

It is important to note that the eigenvalues $\lambda_{m, q}$ are, in general, complex, with the imaginary part representing the frequency of the eigenmode and the real part accounting for the effective losses caused by  intrinsic dissipation and leakage of the mode through the potential barrier of finite width and height. In the presence of the pump, certain modes may formally exhibit negative effective losses (gain) and consequently grow exponentially in time. These modes are responsible for the polariton lasing in the considered geometry. As modes with higher radial quantum numbers $q$ are offset in energy and have substantially higher losses then the ground mode with $q=0$, we neglect them in our further consideration and omit the index $q$ everywhere below. 

Now we can consider the influence of the rotating potential as a small perturbation. 
Projecting Eq.~\eqref{EqGPELinearized} onto the unperturbed solutions of Eq.~\eqref{EqLinearizedUnpertEq}, we obtain the following equation for the amplitude coefficients $C_m$:  
\begin{widetext}
\begin{eqnarray}
\label{EqEqForDecomposeAmplitudes}
    \dot C_{m}=\lambda_{m} C_{m} + (i+\alpha)\left( \eta_{m-} C_{m-\Delta l} e^{i\Delta l \Omega t}+\eta_{m+} C_{m+\Delta l} e^{i\Delta l \Omega t} \right), \label{mode_approx_1}
\end{eqnarray}
\end{widetext}
where $\eta_{m\pm}= 2\pi \int f(r) R_{m\pm\Delta l} R^{*}_{m} rdr$ are overlap integrals that provide the coupling coefficients of the mode $m$ to the modes $m\pm \Delta l$.  

\begin{figure}[tb!]
\begin{center}
\includegraphics[width=\linewidth]{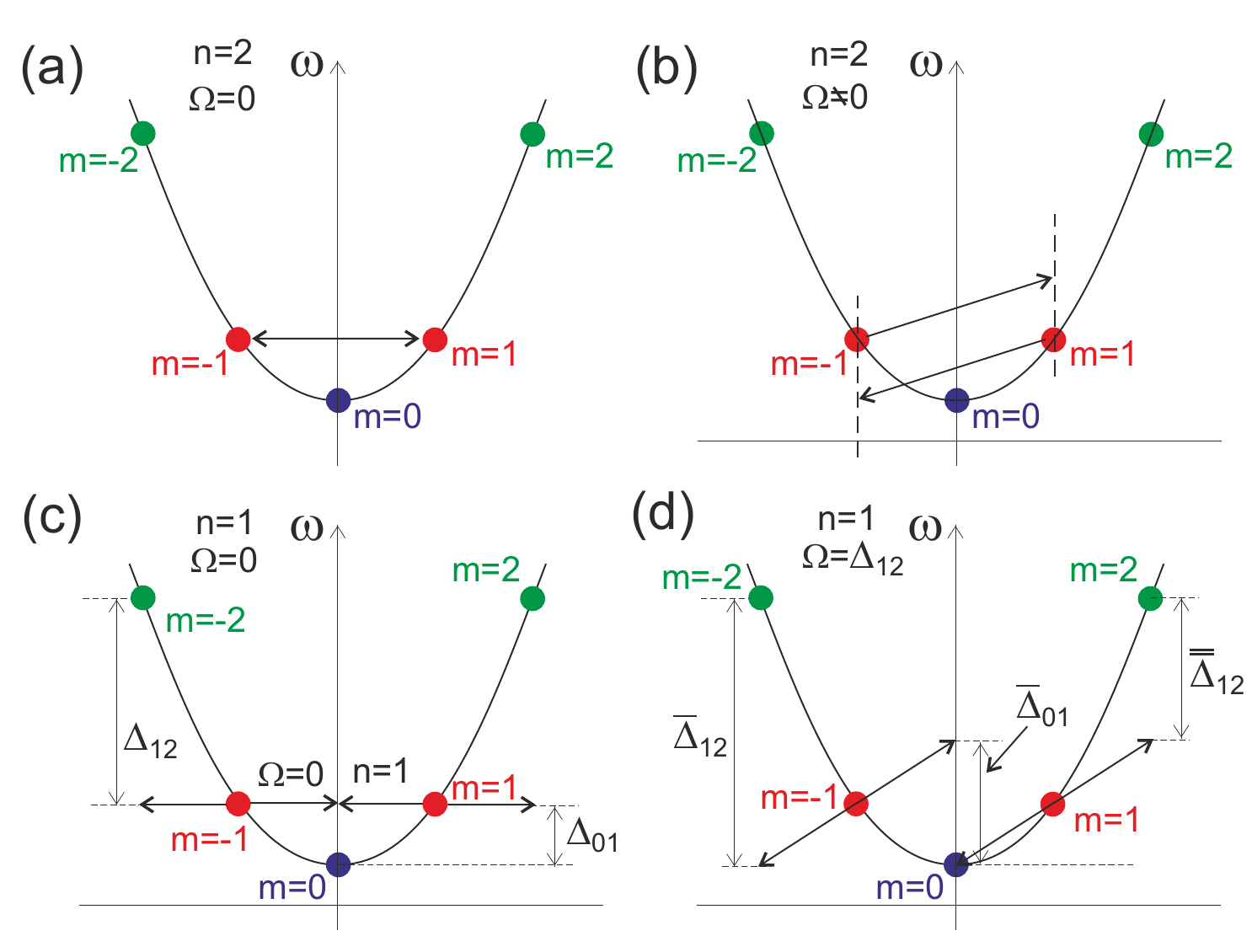}
\end{center}
\caption{ \label{fig1}
(Color online) Schematic representation of energy levels of linear modes of polaritons in a symmetric potential. The vertical axis represents eigenenergies of the modes, the horizontal axis denotes the quantized angular momentum~$m$. Only the modes corresponding to the lowest radial quantum number are displayed. Panel~(a) illustrates the coupling of the states with $m=\pm 1$ by a stationary potential with $\Delta l = 2$. Panel (b) demonstrates the detuning of states from resonance due to the suppression of hybridization of the $m=1$ and $m=-1$ modes when the potential rotates. Panels (c) and (d) illustrate the hybridization of the $m=\pm 1$ states by stationary and rotating potentials with $\Delta l=1$, respectively (see the text for more details).  }
\end{figure}

Let us consider the condition that needs to be satisfied for strong intermode coupling. The energies of the eigenmodes are schematically shown in Fig.~\ref{fig1} where the horizontal axis is the angular index of the mode and the vertical axis is the mode frequency. The condition of the strong coupling of the two modes requires that the angular indices of the modes, $m_{1}$ and $m_{2}$, differ by the angular index of the potential, $ l$, and the frequencies of the interacting states, $\omega _{m1}$ and $\omega _{m2}$, differ by  $\Omega \Delta l$. This can be formulated as a phase matching condition: 
\begin{subequations}
\begin{eqnarray}
&&m_1=m_2 +  l,\\
&&\omega_{m1}=\omega_{m2} + \Omega \Delta l.    
\end{eqnarray}
\end{subequations}

Let us start with the case $\Delta l=2$. One can easily see that for zero angular velocity, stationary potential strongly couples the states with $m=\pm 1$, which has degenerate energies, see Fig.~\ref{fig1}(a). When the potential starts rotating, effective detuning in the rotation frame appears, which weakens the coupling, as it is illustrated in Fig.~\ref{fig1}(b).  It should be noted that in general a rapidly rotating potential can efficiently couple states such as $m=1$ and $m=2$ or $m=1$ and $m=3$. In these cases the physics of the mixing is similar to that between $m=0$ and $m=\pm 1$ states. However, this latter case requires slower rotation velocities to observe the discussed effects, which is advantageous from the experimental viewpoint. Thus in this paper we focus on the rotating potential with $\Delta l=1$, which couples the states with $m=0$ and $m=\pm 1$.

The stationary potential corresponding to $\Delta l = 1 $ cannot efficiently couple the states because the states with $m_2-m_1=1$ always have different eigenfrequencies (see Fig.~\ref{fig1}(c), where it can be observed that the states with $m=\pm 1$ are detuned by some frequency $\Delta_{01}$ from the state with $m=0$ and by $\Delta_{12}$ from the states with $m=\pm 2$). However rotation with velocity $\Omega=\Delta_{01}$ couples the state $m=1$ to the state $m=0$, see Fig.~\ref{fig1}(d). It is worth mentioning that this rotation is not sufficient to couple the state $m=1$ to the state $m=2$ as non-zero detuning $\bar{\bar{\Delta}}_{12}$ is maintained. Note, that this rotation couples the state $m=-1$ neither to the state $m=0$ nor to the state $m=-2$ as corresponding detunings in the rotating frame are increased to $\bar \Delta_{01}$ and $\bar \Delta_{12}$, respectively. This way the symmetry $\theta \rightarrow -\theta$ gets broken: when the rotating potential is present, the state with angular momentum $m=1$ undergoes hybridization with the state $m=0$, while the state with angular index $m=-1$ remains almost unaffected by rotating potential. Therefore one can anticipate that the properties of the states $m=\pm 1$ change differently when subjected to a rotating potential.

\section{Numerical studies of polaritons in rotating potentials with angular index $\Delta \mathit{l}=1$}

\begin{figure}[tb!]
\begin{center}
\includegraphics[width=\linewidth]{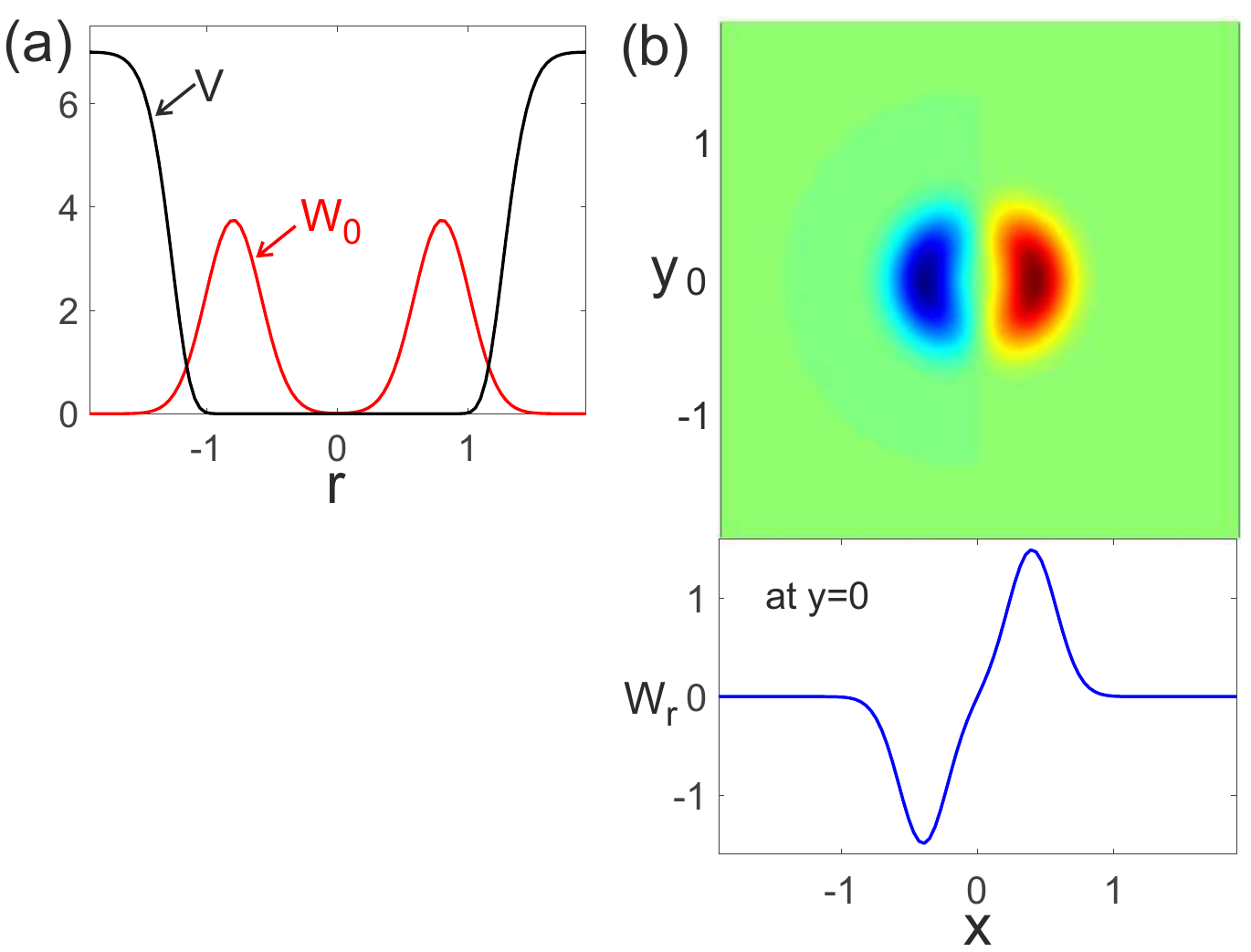}
\end{center}
\caption{ \label{PT_scalar_1}
(Color online) (a) Potentials created by the microstructuring, $V$, and by the radially symmetric pump,~$W_0$. (b)~The spatial distribution of the rotating potential~$W_r$. The upper part of the panel shows 2D pattern of the potential, the cross section of the potential $W_r$ by $y=0$ plane is shown in the lower part of the panel. In numerical simulations potential $V$  is defined as $V=V_0\left\{\exp[ -(r-R_{V})^8/w_V^8]+\exp[ -(r+R_{V})^8/w_V^8]\right\}$, with $V_0=7$, $R_V=2.25$ and $w_V=1$. The radially symmetric potential is $W_0=\tilde W_{0}\left\{\exp[ -(r-R_{W})^2/w_W^2]+\exp[ -(r+R_{W})^2/w_W^2]\right\}$ with $\tilde W_0=3.75$, $R_W=0.8$ and $w_W=0.3$ and, finally, the rotating potential $W_r=\tilde W_{r}\left\{\exp[ -(r-R_{r})^2/w_r^2]-\exp[ -(r+R_{r})^2/w_r^2]\right\}\cos(\theta)$ with $\tilde W_r=1.5$, $R_r=0.4$ and $w_r=0.25$.}
\end{figure}

To check our hypothesis, we performed numerical simulations using potentials shown in Fig.~\ref{PT_scalar_1}. The dimensionless parameter $\alpha$ defining the ration of the gain to the polariton frequency shift created by the pump was set to~0.33. The dimensionless nonlinear coefficients accounting for the polariton interactions and nonlinear losses were chosen to be $h=0.0186$ and $\tilde h=0.001$, respectively.
The corresponding values of dimensional parameters in Eq.~\eqref{master_eqs} are given in~\cite{RealParamValues}.

It has been established~\cite{yulin2016spontaneous} that by adjusting the parameters of a pump beam, it is possible to selectively excite the modes with well defined angular and radial indices. For our simulation, we choose the potential $W_0$ such that in the absence of the rotating pump a condensate with lowest radial index and angular momentum $\pm 1$ is created. Because of the symmetry, the modes with $m=+1$ and $m=-1$ are excited with equal probability. The potential also supports the fundamental mode, which exhibits slightly higher effective losses compared to the modes with $m=\pm 1$ (given the chosen  set of parameters). The other modes have poor localization and high effective losses, which allows us safely to assume that they do not significantly impact the dynamics.

\begin{figure}[tb!]
\begin{center}
\includegraphics[width=\linewidth]{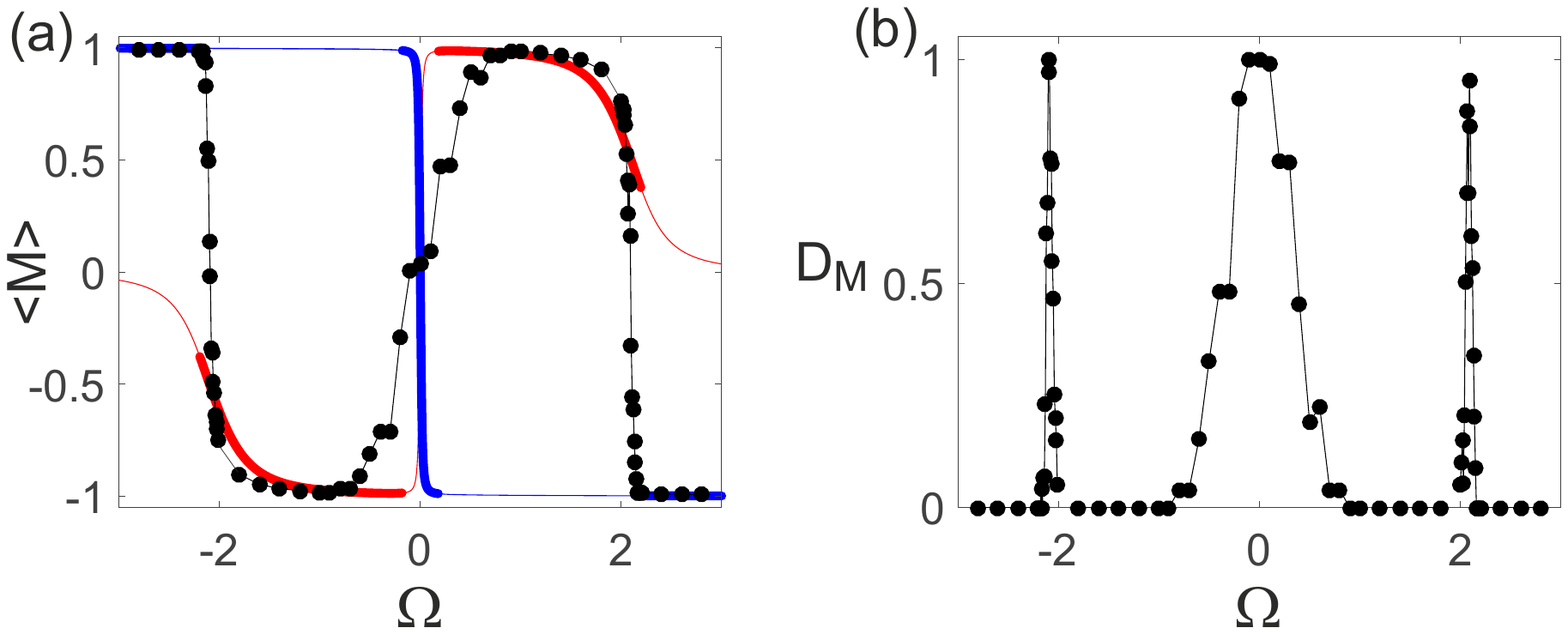}
\end{center}
\caption{ \label{fig3_new}
(Color online)
(a) The normalized angular momenta of stationary polariton states obtained from 2D numerical simulations (black dots) and of the two fastest growing modes obtained using the perturbation theory (red and blue curves) as functions of the potential rotation velocity~$\Omega$.
The thicker sections of the curves indicate the mode with the highest growth rate.
In the 2D simulations, the angular momenta are averaged over the results of the series of $N=100$ runs, each starting from random initial conditions, $<M>=\frac{\sum_{j =1}^N M_j}{N}$ where $M_j$ are the normalized angular moment by $j$-th run.
(b) The dispersion $D_M=\frac{\sum_{j=1}^N (M_j-<M>)^2}{\sum_{j=1}^N M_j^2}$ of the angular momenta obtained from 2D simulations, see text for more details.
The thin line connecting the circles is the guide for eyes.  
}
\end{figure}

We performed a series of numerical simulations corresponding to different angular velocities of the rotating potential ($N=100$ simulations for each frequency). We calculated the number of polaritons 
$E=\int |\psi|^2 d^2\mathbf{r}$ 
and the angular momentum 
\begin{equation}
\cM=i \int \psi^{*}(y \partial_x -x\partial_y)\psi d^2\mathbf{r}    
\end{equation} 
of stationary polariton states in each round of the simulations. To characterize the rotation of the condensate, it is convenient to introduce its normalized angular momentum defined as $M=\cM/E$. The dependency of this quantity averaged over the number of realizations on angular velocity of the potential is shown in Fig.~\ref{fig3_new}(a) by solid circles. Panel (b) shows the dispersion of the angular momentum defined as:
\begin{equation}
D_M=\frac{\sum_{j=1}^N (M_j-<M>)^2}{\sum_{j=1}^N M_j^2},    
\end{equation}
where $M_j$ is the normalized angular momentum calculated in $j$-th round of simulations, and ${<M>}=\left(\sum_{j=1}^N M_j\right)/N$ is normalized angular momentum averaged over $N$ simulations with different random initial conditions. It is seen that the dispersion deviates significantly from zero only in the vicinity of the  potential angular velocities at which a change in the direction of polariton rotation occurs. A comprehensive explanation of this fact is provided in the subsequent section, where the perturbation theory is developed.

\begin{figure}[tb!]
\begin{center}
\includegraphics[width=.9\linewidth]{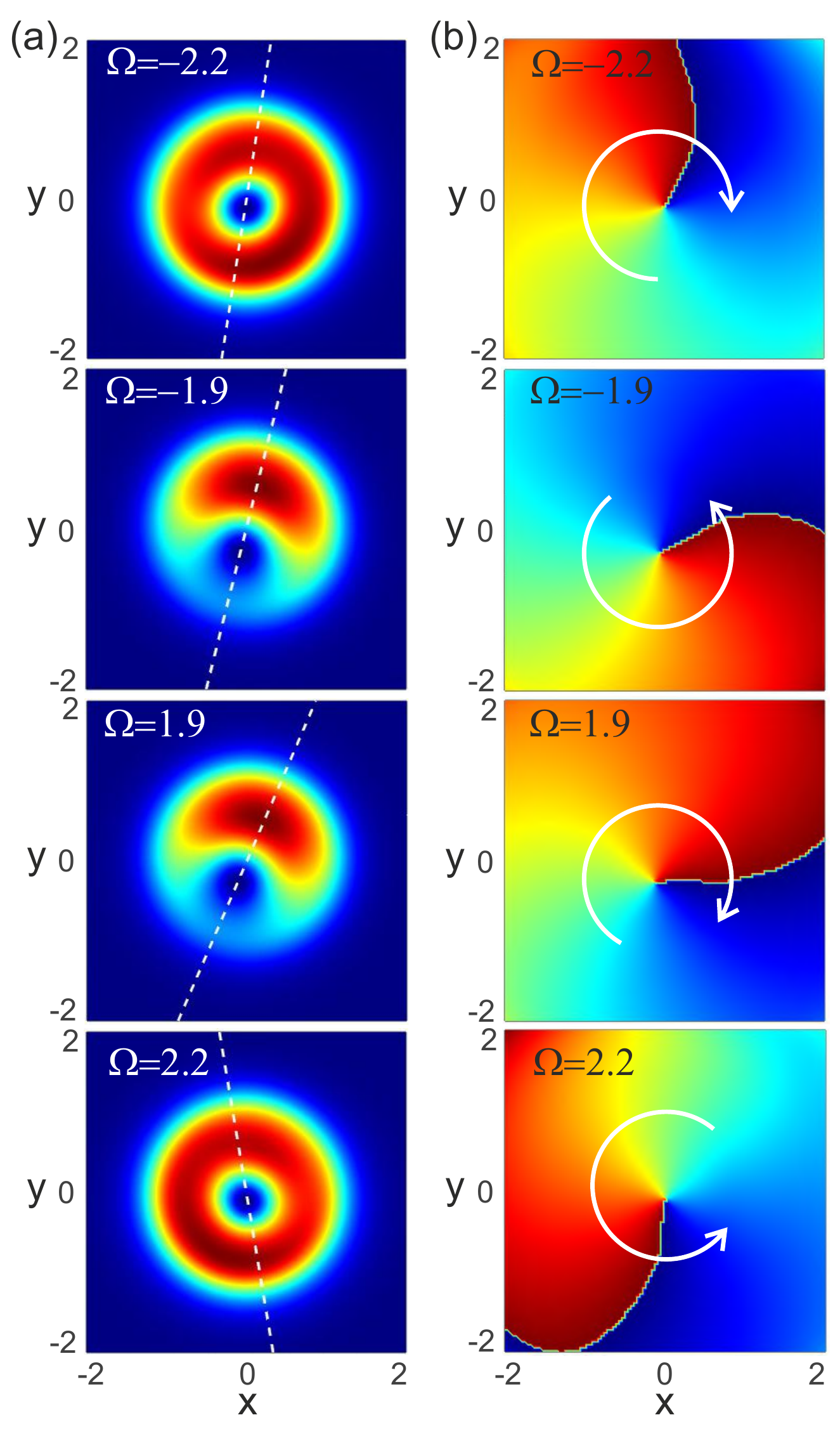}
\end{center}
\caption{ \label{fig4_new}
(Color online)
The density (a) and phase (b) distributions  for the stationary polariton states for the case of a potential corresponding to $\Delta l=1$ rotating at angular velocities $\Omega=-2.2$, $\Omega=-1.9$, $\Omega=1.9$ and $\Omega=2.2$. The thin blue dashed lines in panels (a) show the in-plane symmetry axis of the rotating potential. White round arrows in panel (b) indicate the direction of the rotation of a condensate. 
It is seen in (b) that the phase gradient is directed oppositely for the angular velocities of the potential equal to $\Omega=1.9$ and $\Omega=2.2$ (the same is true for the $\Omega=-1.9$ and $\Omega=-2.2$ ). This means that the polaritons can rotate in the same and in the opposite directions as the potential.  This should also be mentioned the polariton density distributions are quite different at $\Omega=1.9$ and $\Omega=2.2$. At $\Omega=1.9$ the polaritons look like a pulse going round the trap whereas at $\Omega=2.2$ the polariton flux and density are distributed more or less evenly along the angular coordinate. 
}
\end{figure}

The distinction between the states formed at different rotation velocities is illustrated in Fig.~\ref{fig4_new}, which displays the density and phase distribution of the stationary polariton states calculated at $\Omega = \pm 2.2$ and $\Omega = \pm 1.9$. 
From this visualization, it can be deduced that polaritons generated in potentials rotating at velocities $\Omega= \pm 1.9$ and $\Omega= \pm 2.2$ exhibit opposite rotation directions.

The results of the simulation clearly demonstrate that the normalized angular momentum of the stationary polaritons is an odd function of the angular velocity of the potential: $M(\Omega)=-M(-\Omega)$. However, a significant finding is that there exist specific non-zero potential rotation velocities at which both the sign and the absolute value of the polariton angular momentum  undergo abrupt changes. Below we show that these velocities act as thresholds that separate different ranges of potential angular velocities, where the polaritons condense into distinct states. Importantly, these stationary states possess the rotation velocities with different signs and absolute values. Figure~\ref{fig3_new}  clearly illustrates this transition occurring at~$\Omega \approx \pm 2$.

\section{Perturbation theory  for polaritons in rotating potentials with $\Delta l=1$ }

To explain the observed phenomenon, we construct a simple coupled mode theory based on Eq.~\eqref{mode_approx_1}. In our case, when $\Delta l=1$, we take into account the interaction between the modes with $m=0$ and $m=\pm 1$ only assuming that the coupling with other modes is negligible within the frequency range we consider. First, we determine the parameters of the eigenmodes for the chosen stationary potentials $V$ and $W_0$ in the absence of the rotating potential $W_r=0$. The analysis of the field evolution reveals the presence of a few localized modes in the spectrum, with only three modes exhibiting relatively high Q factors: ${m = \pm 1}$ and $m = 0$. Additionally, the frequency separation between the modes with ${m = 0}$ and ${m = \pm 1}$ was found to be approximately $2$ in the dimensionless units used by us, while the frequency difference between the other modes was considerably larger, see Appendix for more details. These findings support our assumption that a three-mode approximation is sufficient to analyze the dynamics of the system. 

Let us rewrite Eq.~\eqref{EqEqForDecomposeAmplitudes} in the three modes approximation:
\begin{widetext}
\begin{subequations}
\begin{eqnarray}
&&\dot C_{-1}=(-\gamma_1 + i\omega_1) C_{-1} + (i+\alpha) \eta_{10} C_{0} e^{i \Omega t}, \label{mode_approx_2_1} \\
&&\dot C_{0}=(-\gamma_0 + i\omega_0) C_{0} + (i+\alpha)\left( \eta_{10} C_{1} e^{i \Omega t}+\eta_{10} C_{-1} e^{-i \Omega t} \right), \label{mode_approx_2_2} \\
&&\dot C_{1}=(-\gamma_1 + i\omega_1) C_{1} + (i+\alpha)\eta_{10} C_{0} e^{-i \Omega t},  \label{mode_approx_2_3}
\end{eqnarray}
\end{subequations}
\end{widetext}
where $\gamma_1$ and $\gamma_0$ are the effective losses of the modes with the angular indices $m=\pm 1$ and $m=0$, respectively, which take into account the radiative losses and the effect of the non-rotating pump, $\omega_1$ and $\omega_0$ are real eigenfrequencies of the modes, $\eta_{10}$ is the interaction strength between the modes. These parameters were determined from  2D numerical modelling (see Appendix for more details).  

We set the frequency $\omega _0$ of the mode with $m = 0$ as a reference frequency and express the other frequencies as detunings from~$\omega _0$. We eliminate the explicit time dependence of the coupling coefficients by introducing new complex amplitudes 
\begin{subequations}
\begin{eqnarray}
&&A_1=C_1 \exp^{-i(\omega_0 +\Omega)t},\\
&&A_0=C_0 \exp^{-i\omega_0 t},\\
&&A_{-1}=C_{-1} \exp^{-i(\omega_0 -\Omega)t} .  
\end{eqnarray}
\end{subequations}

The equations for $A_{0, \pm 1}$ can be written in the following form:
\begin{subequations}
\begin{eqnarray}
&&\dot A_{-1}=(-\gamma_1 + i\Delta_1+i\Omega) A_{-1} + (i+\alpha) \eta_{10} A_{0} ,  \quad\label{mode_approx_3_1} \\
&&\dot A_{0}=-\gamma_0 A_{0} + (i+\alpha)\eta_{10}\left(  A_{1} + A_{-1} \right),  \quad\label{mode_approx_3_2} \\
&&\dot A_{1}=(-\gamma_1 +i\Delta_1 - i\Omega) A_{1} + (i+\alpha)\eta_{10} A_{0},   \quad \label{mode_approx_3_3}
\end{eqnarray}
\end{subequations}
where $\Delta _1=\omega_1-\omega_0$ is the detuning between the frequencies of the modes with $m=\pm 1$ and $m=0$. 

\begin{figure}[tb!!!]
\begin{center}
\includegraphics[width=\linewidth]{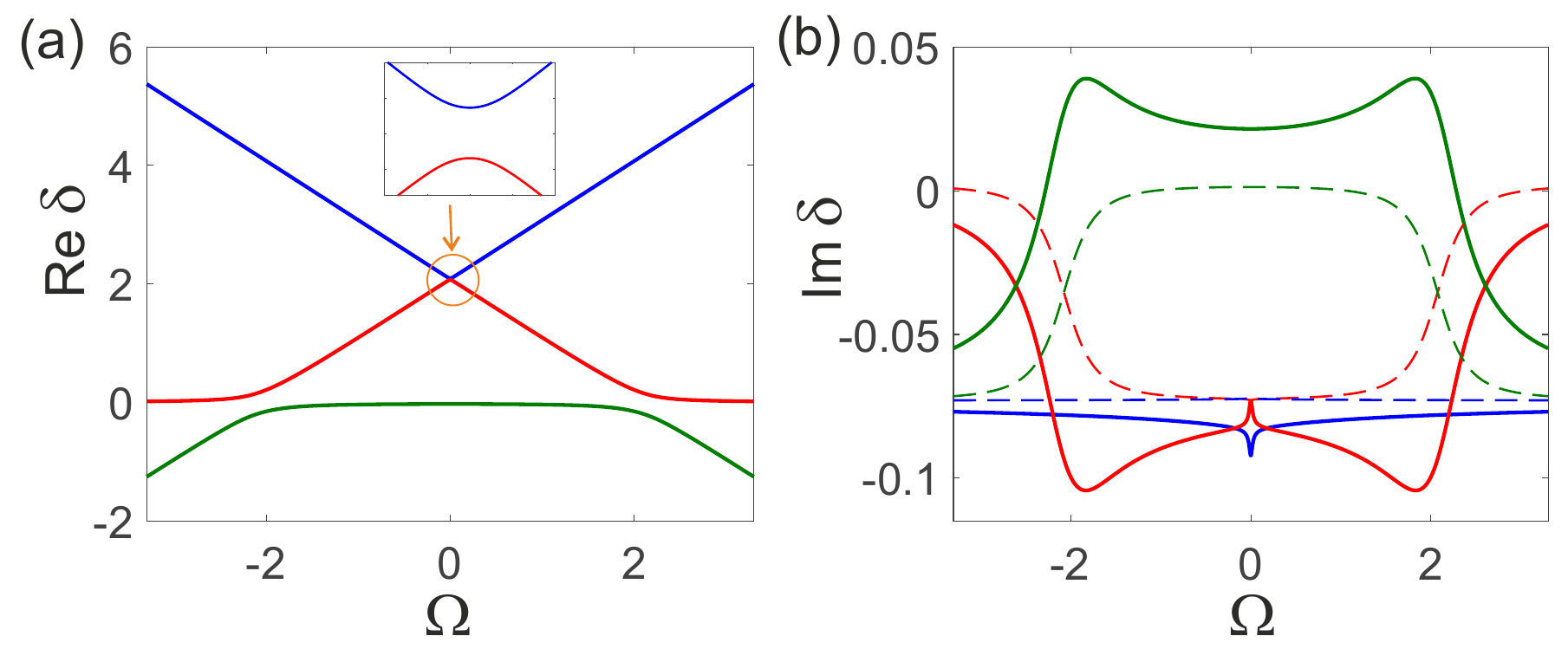}
\end{center}
\caption{\label{figDisp_new}
(Color online)
(a) The dependencies of the real part of the eigenfrequencies of the modes calculated in framework of the coupled mode theory as the functions of the potential rotation velocity $\Omega$ (b) The dependencies of the imaginary part of the eigenfrequencies of the modes calculated in framework of the coupled mode theory as the functions of the potential rotation velocity $\Omega$. The dashed lines illustrate the case where the rotating potential is pure conservative. Negative values of the imaginary part correspond to the modes growing in time.}
\end{figure}

Looking for the solution in the form $A \sim \exp(i \delta t)$, it is easy to find the eigenfrequencies $\delta$ and the eigenvectors $\vec A = (A_{-1}, A_0, A_{1})^T $ of the modes as functions of the angular velocity~$\Omega$ of the potential. The dependencies of the real and imaginary parts of the eigenfrequencies $\delta$ are shown in Fig.~\ref{figDisp_new} for the parameters corresponding to 2D simulations discussed above. It should be noted that because of the pump, the values of $\gamma_{0, \pm1}$  can be negative, indicating that corresponding modes in the unperturbed problem (without the rotating potential) experience linear gain and thus grow in time.

As shown in Fig.~\ref{figDisp_new}(a), the real parts of the eigenfrequencies $\delta$ exhibit gaps at the rotation velocities $\Omega=0$ and $\Omega = \pm \Delta_1$.  The gap at $\Omega=0$ arises from the interaction of the modes with $m=\pm 1$ mediated by the mode with $m=0$. Consequently, this gap is quadratic in $\eta_{10}$ and is thus small for shallow rotating potentials. The gaps at $\Omega = \pm \Delta_1$ occur due to the resonant scattering between the modes $m=0$ and $m=1$ caused by the rotating potential or, for opposite sign of $\Omega$, between the modes with $m=0$ and $m=-1$.

It is worth noting that in the vicinity of the resonant rotation velocity $\Omega \approx \pm \Delta _1$ the effect of the third mode can be neglected, and the eigenfrequencies of two hybrid modes read:
\begin{multline}
\delta_{\pm}=\frac{1}{2} \left[i(\gamma_0 +\gamma_1)+\Delta_1+\Omega  \right. \\
\left. \pm i \sqrt{(\gamma_0-\gamma_1+i(\Delta_1+\Omega))^2+4(1-i\alpha)^2\eta_{10}^2} ,   \label{disp1_1} \right]
\end{multline}
for $\Omega \approx -\Delta_1$ and 
\begin{multline}
\delta_{\pm}=\frac{1}{2} \left[ i(\gamma_0 +\gamma_1)+\Delta_1-\Omega \right. \\ 
\left. \pm i \sqrt{(\gamma_0-\gamma_1+i(\Delta_1-\Omega))^2+4(1-i\alpha)^2\eta_{10}^2} \right]   \label{disp1_2}
\end{multline}
for $\Omega \approx \Delta_1$.
Let us remark that approximate analytical expression for the dependencies of $\delta(\Omega)$ can be obtained for the gap at $\Omega=0$ as well.  

It is important that the eigenfrequencies are complex with their imaginary parts governing the growth (for $\text{Im} \, \delta <0$) or decay (for $\text{Im} \, \delta >0$) of a mode. As it can be observed in Fig.~\ref{figDisp_new}(b), to the left of the resonance $\Omega=\Delta_1$, the red-colored mode is the fastest growing one, while the blue-colored mode is the second fastest growing. To the right of the resonance, the roles of the modes reverse, with the blue mode becoming the fastest growing. 

Based on the linear mode analysis, we can propose the following scenario for the crossovers observed in our numerical simulations. If initial conditions are taken in the form of very low intensity noise, the fastest growing mode eventually dominates and suppresses the growth of the other modes once it reaches the nonlinear regime. This scenario takes place when the lasing threshold is surpassed by one mode only. 

When the pump intensity is close to the lasing threshold, nonlinear effects lead to the saturation of the mode but do not significantly alter its structure. In this case, one can anticipate that the structure of the stationary state closely resembles that of the fastest growing linear mode, and their normalized angular momenta coincide.   

The dependencies of the normalized angular momenta~$M$ of the fastest and the second fastest growing linear modes are shown in Fig.~\ref{fig3_new} in blue and red, respectively. The thicker lines indicate the range of $\Omega$ where a mode is the fastest growing. It is clearly seen that the angular momenta of the stationary states observed in the full scale 2D simulations are very close to the angular momenta of the fastest growing linear mode. 

Thus our perturbation theory shows that if the pump is close to the lasing threshold, the stationary polariton state is determined by the fastest growing linear mode. Furthermore, the swap of the fastest and the second fastest growing modes causes the abrupt change of the stationary polariton state. 
Now, let us delve into the reason behind the substantial deviation of the dispersion of the normalized angular momentum, as observed in a series of 2D simulations, in the vicinity of this transition, see Fig~\ref{fig3_new}.

The reason for this is that a mode having smaller increment can win the competition if its initial amplitude is sufficiently higher than that of the fastest growing mode. To estimate the ratio of initial amplitudes at which the second fastest growing mode can prevail, we make a simple analysis. Let us assume that the nonlinear effects come to play, and the mode start suppressing its competitors when its amplitude becomes equal to $a_{th}$. We denote the increment and the initial amplitude of the fastest growing mode as $\gamma_{f}$ and $a_{f}$. Similarly, the increment and the amplitude of the second fastest growing mode are $\gamma_{s}$ and $a_{s}$. Let us find the condition providing that both modes reach the value $a_{th}$ at the same time $a_{f}\exp(\gamma_{f} t)=a_{s}\exp(\gamma_{s} t)=a_{th}$. This time is given by $\ln({a_{th}}/{a_{f}})$. The second fastest growing mode reaches the critical value earlier than the fastest mode if  $a_{s}>a_{f}\exp \left(\frac{\gamma_{f}-\gamma_{s}}{\gamma_{f}} \ln \frac{a_{th}}{a_{f}} \right)$. 

The latter formula reveals that the probability of the second fastest growing mode to win the competition depends on the intensity of the noise taken as the initial conditions. However, if the initial noise is weak, $a_f, a_s \ll a_{th}$, the fastest and the second fastest growing modes have a comparable probability of winning only if the difference of the increments $\gamma_f-\gamma_s$ is small. This occurs at the velocities close to the critical velocity  at which the fastest and the second fastest modes swap, as at this point the increments are equal. This explains why the dispersion of the angular momentum in Fig.~\ref{fig3_new} is large in the vicinity of resonant angular velocity $\Omega \approx 2$. In the regions of $\Omega$ where the difference in increments is large, the fastest growing mode always emerges as the winner. 

It should be noted that this crossover does not occur in the case of purely conservative rotating potential. Indeed, for this case there is no swap of the fastest and the second fastest growing modes, as it can be seen in Fig.~\ref{fig4_new}(b). Hence, one should expect smooth dependency of the average angular momentum of polaritons on the rotation velocity of the potential as it is shown by the blue line in Fig.~\ref{fig3_new}.

Now, let us briefly discuss the behavior at small rotation velocities. In this case there is another swap between the fastest and the second fastest growing modes for small angular velocities of the rotating potential around $\Omega\approx \pm 0.2$ for the chosen parameters. Consequently, there exists another threshold velocity $\Omega$ at which the normalized angular momentum of the condensate changes. However, the difference of the growth rates of the competing modes is relatively small, and to observe the crossover, long simulation times are required, which renders full-scale 2D simulations costly.

\begin{figure}[tb!]
\begin{center}
\includegraphics[width=\linewidth]{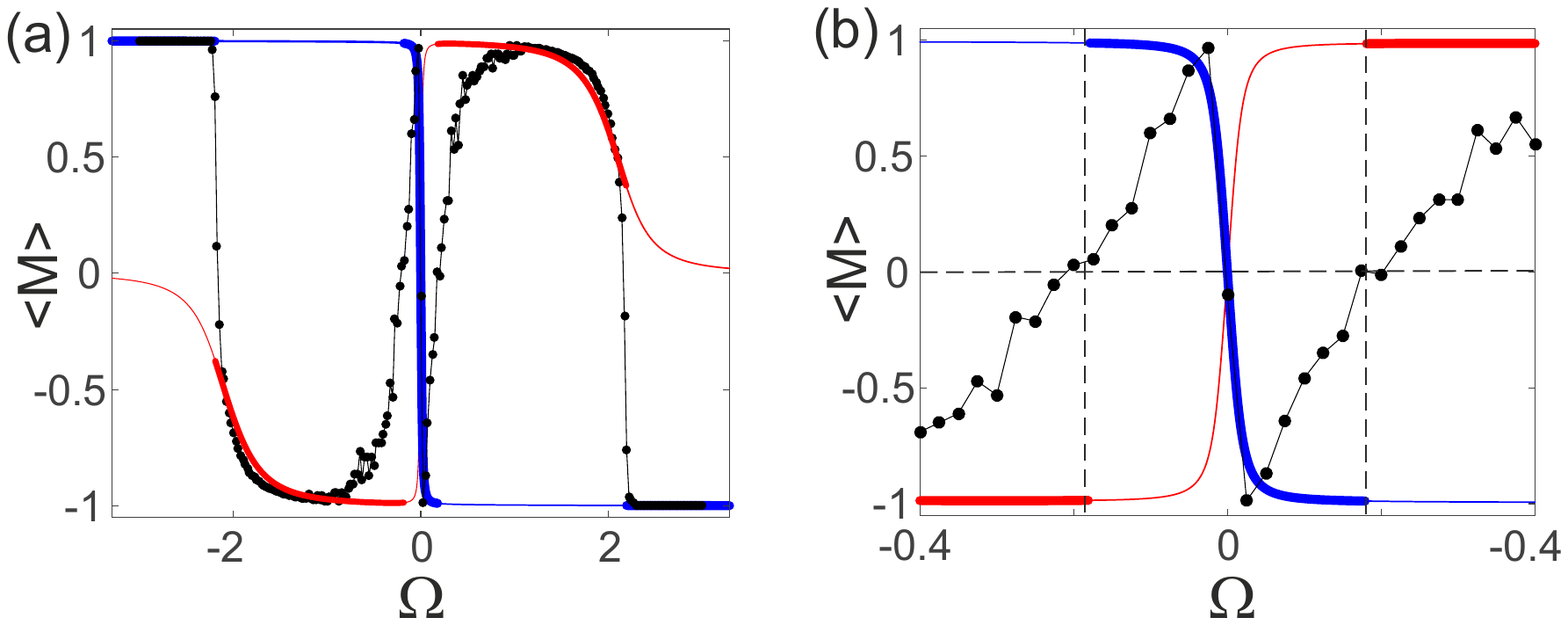}
\end{center}
\caption{ \label{fig5_new}
(Color online)
(a) Normalized angular momenta of stationary states obtained from the coupled mode theory are shown by black filled circles (the thin black line connecting the circles is the guide for eye). The red and  blue lines show the normalized angular momenta of the two fastest growing modes as functions of the potential rotation velocity~$\Omega$. The thicker parts of the curves mark fastest growing modes. (b) Same as in panel (a), but for narrower frequency range. The vertical dashed lines mark the swap at of the fastest and the second fastest growing modes. The average normalized angular momentum changes it sign at  $\Omega=0$ and $\Omega \approx \pm 0.2$ $\Omega \approx \pm 2.$.    }
\end{figure}

For this reason we performed numerical simulations of the coupled mode approximation generalized for the weakly nonlinear case. The dynamic equations for the modes can then be written as follows:
\begin{widetext}
\begin{subequations}
\begin{eqnarray}
&&\dot A_{-1}=(-\gamma_1 + i\Delta_1+i\Omega) A_{-1} + (i+\alpha) \eta_{10} A_{0}+\left(\epsilon_{01} |A_{0}|^2+\epsilon_{11} |A_{-1}|^2+2\epsilon_{11} |A_{1}|^2   \right)A_{-1} , \label{mode_approx_3_1NL} \\
&&\dot A_{0}=-\gamma_0 A_{0} + (i+\alpha)\eta_{10}\left(  A_{1} + A_{-1} \right) +\left( \epsilon_0 |A_{0}|^2+\epsilon_{01} |A_{1}|^2+\epsilon_{01} |A_{-1}|^2   \right)A_0 , \label{mode_approx_3_2NL} \\
&&\dot A_{1}=(-\gamma_1 +i\Delta_1 - i\Omega) A_{1} + (i+\alpha)\eta_{10} A_{0}+\left(\epsilon_{01} |A_{0}|^2+\epsilon_{11} |A_{1}|^2+2\epsilon_{11} |A_{-1}|^2   \right)A_{1},   \label{mode_approx_3_3NL}
\end{eqnarray}
\end{subequations}
\end{widetext}
where  $\epsilon_0$, $\epsilon_{ij}$ are complex constants defining the intra-mode ($\epsilon_{0}$, $\epsilon_{11}$) and inter-mode interactions ($\epsilon_{01}$) (see Appendix for more details). The results are presented in Fig.~\ref{fig5_new}. 

The results obtained from the coupled mode theory confirm the conclusion that the normalized angular momentum of polaritons near the lasing threshold is determined by the fastest growing linear mode. Figure~\ref{fig5_new}(a) illustrates that the averaged angular momenta of the stationary polaritons closely align with the normalized angular momenta of the fastest growing linear mode, except in narrow ranges of the angular velocity $\Omega$ where the transitions between different polariton states occur. 

Numerical simulations of a coupled mode equations are much faster and allow for the collection of statistics through a large number of runs using very small random fields as initial conditions. This enables the accurate resolution of the transition occurring at $\Omega \approx 0.2$, as well as of the change of the polariton rotation at $\Omega=0$. Figure~\ref{fig5_new}(b) shows that the dependency of the normalized angular momentum of the stationary polaritons on $\Omega$ passes through zero at $\Omega=0$ and at $\Omega \approx \pm 0.2$.

It should be noted that the smoothness of the transitions at $\Omega \approx 0.2$  is determined by the intensity of the initial noise used as the initial conditions. In an ideal scenario, these transitions should be sharp. However, achieving the desired sharpness is challenging due to the requirement of extremely low intensities for the initial conditions, necessitating very long numerical simulations. This is why in our full-scale 2D simulations, we were unable to resolve these transitions completely, but instead observed a large dispersion of the normalized angular momentum at the indicated angular velocities $\Omega$, as shown in Fig.~\ref{fig3_new}(b). This means that the difference of the increments of the fastest and the second fastest modes is so small that, for the chosen statistical properties of the initial conditions, these modes have comparable probabilities of winning.   

At $\Omega = 0$, there is another change in the sign of the angular momentum of the condensate. This change is connected with the symmetry of the fastest growing mode. At $\Omega=0$, the angular momentum of the fastest growing mode is exactly zero. However, the rotation of the potential breaks this symmetry, resulting in the appearance of the angular momentum of the stationary polariton state. Therefore, we can conclude that the changes in the sign of the angular momentum at $\Omega=0$ and $\Omega=\Omega_r$ occur due to different physical reasons. At $\Omega =\Omega_r$, it is due to the swap between the two fastest growing modes, while at $\Omega=0$, it is due to the change in symmetry of the fastest growing mode. 

\section{Comparison of the polariton dynamics in rotating potentials with $\Delta l = 1$ and $\Delta l = 2$}

\begin{figure}[tb!]
\begin{center}
\includegraphics[width=\linewidth]{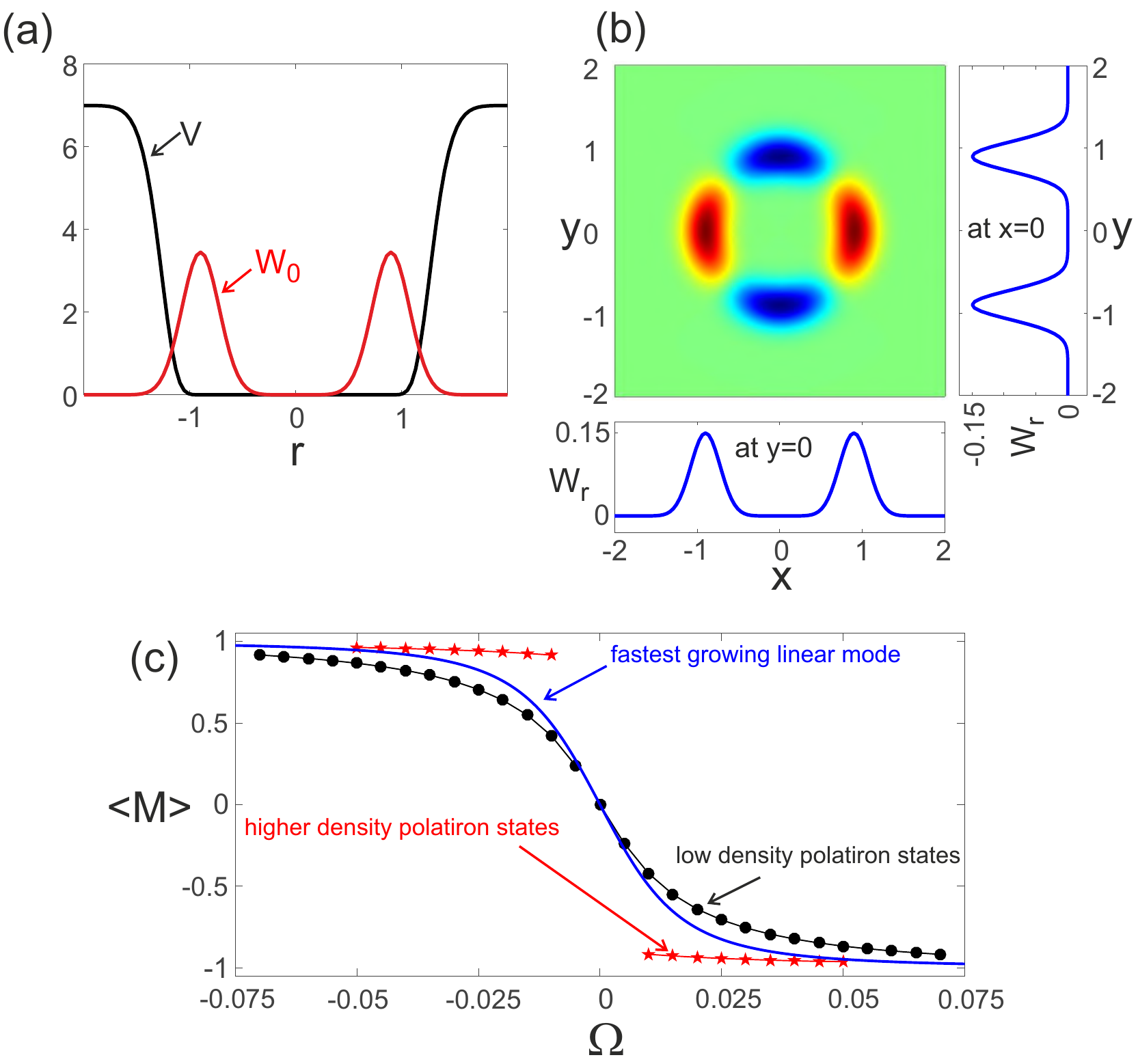}
\end{center}
\caption{ \label{fig6_new}
(a) Potentials created by the microstructuring, $V$, and by the radially symmetric pump,~$W_0$. (b)~The spatial distribution of the rotating potential~$W_r$. The 2D distribution of the potential is shown in the upper left part of the panel,  the cross sections of the potential $W_r$ by $y=0$ and $x=0$ planes are shown in the lower and the upper right parts of the panel correspondingly. The potential is created by the incoherent pump with the angular index difference $\Delta l =2$.
The potential $V$ in numerical simulations are defined as $V=V_0\left\{\exp[ -(r-R_{V})^8/w_V^8]+\exp[ -(r+R_{V})^8/w_V^8]\right\}$, with ${V_0=7}$, ${R_V=2.25}$ and ${w_V=1}$.
The azimuth symmetric potential is $W_0=\tilde W_{0}\left\{ \exp[ -(r-R_{W})^2/w_W^2]+\exp[ -(r+R_{W})^2/w_W^2]\right\}$ with ${\tilde W_0=3.45}$, ${R_W=0.9}$ and ${w_W=0.25}$.
The rotating potential is $W_r=\tilde W_{r}\left\{\exp[ -(r-R_{r})^2/w_r^2]-\exp[ -(r+R_{r})^2/w_r^2]\right\}\cos(2\theta)$ with ${\tilde W_r=0.15}$, ${R_r=0.9}$ and ${w_r=0.25}$.
The averaged (over $100$ simulations starting from a weak noise taken as the initial conditions) normalized polariton angular momentum as a function of the rotation velocity of the potential $\Omega$ is shown in panel (c) by solid black circles. Red circles show the same but for the stronger pump $W_0=3.75$ corresponding to the polariton number 
$E$ in the stationary state approximately $5$ times as much as that for the lower pump with $W_0=3.45$. The blue line shows the dependency of the normalized angular momentum on $\Omega$ for the fastest growing linear mode calculated withing the coupled mode approach for the parameters fitted to the full scale 2D simulations. 
}
\end{figure}

The latter scenario discussed plays a key role in determining the angular momentum of polariton states in potentials that couple states with $m=\pm 1$~\cite{lagoudakis2022}. We performed  numerical simulations using the potential shown in Fig.~\ref{fig6_new}(a) with $\Delta l=2$, that couples modes with $m=\pm1$. The non-rotating axially symmetric potentials are shown in panel~(b). The averaged normalized momenta of the stationary polariton states are shown in Fig.~\ref{fig6_new}(c) for the excitation slightly above the polariton lasing threshold. It is clearly seen that the direction of the rotation of the potential determines the direction of the rotation of the condensate.  

\begin{figure}[tb!]
\begin{center}
\includegraphics[width=\linewidth]{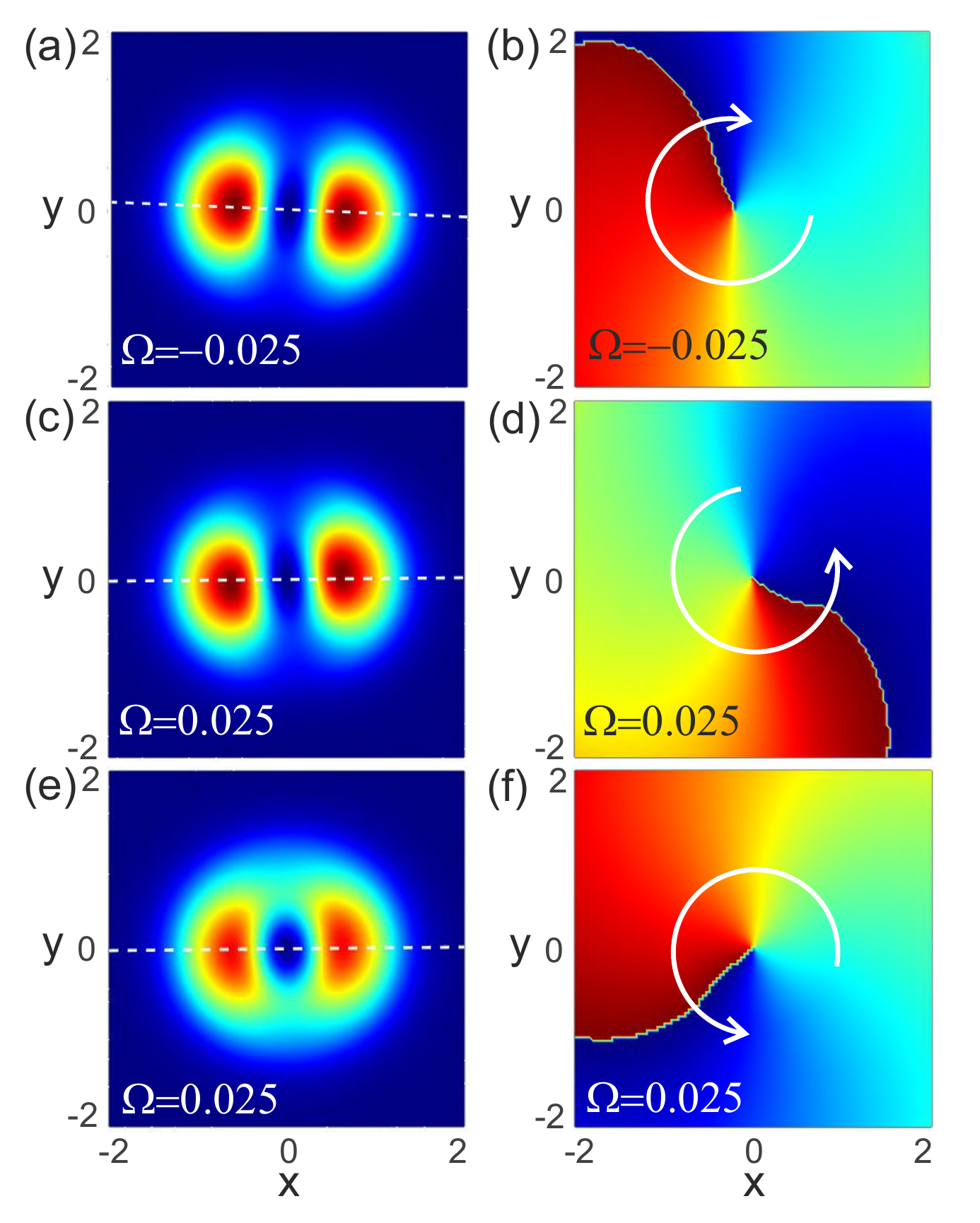}
\end{center}
\caption{ \label{fig7_new}
(Color online)
Density {(a,c,e)} and phase {(b,d,f)} distributions corresponding to the stationary states for the case of the rotating potential with $\Delta l=2$ and rotation velocities $\Omega=-0.025$ and  $\Omega=0.025$. Thin white dashed lines in the left panels show the in-plane rotating potential symmetry axis passing through the potential maxima.
Arrows in the right panels indicate rotation direction of the polariton states.
For (a--d), the stationary pump slightly exceeds the lasing threshold, $W_0=3.45$, whereas for (e,f), the pump is more intense, $W_0=3.75$. The increase of the pump from $W_0=3.45$ to $W_0=3.75$ results in the growth of the total polariton number $E$ in the stationary states by approximately $5$ times. It is seen that the for the higher incoherent pump the variation of the polariton density along the angular coordinate becomes less pronounced.
}
\end{figure}

The density and the phase distributions of the stationary polariton states are shown in Fig.~\ref{fig7_new} (a,c,f) and Fig.~\ref{fig7_new} (b,d,e), respectively. The panels (a--d) correspond to the case of the pump slightly exceeding the lasing threshold. The direction of the condensate rotation is shown by  white arrows. Notably, the polariton density distribution consists of two lobes separated by deeps. The separation becomes more pronounced at low rotation velocities, see panels~(a,c).  

To establish the connection between the angular momentum of a stationary state and that of the corresponding fastest growing mode, we examined the coupled mode model~\cite{lagoudakis2022}. In the case of a potential with ${\Delta l=2}$, it is sufficient to take into account only two modes with ${m = \pm 1}$. By substituting
\begin{equation}
A_{\pm1}=C_{\pm 1} e^{ -i(\omega_1 \pm 2\Omega)t }  
\end{equation} 
into~(\ref{mode_approx_1}), we get:
\begin{eqnarray}
    \dot A_{\pm 1}=-(\gamma_1 \pm 2 i\Omega) A_{\pm 1} + (i+\alpha) \eta_{\pm 1} A_{\mp 1}, \label{mode_approx_LGD_CMA} 
\end{eqnarray}
where $\eta_{\pm 1}$ is the coupling strength between the modes. 

\begin{figure}[tb!]
\begin{center}
\includegraphics[width=\linewidth]{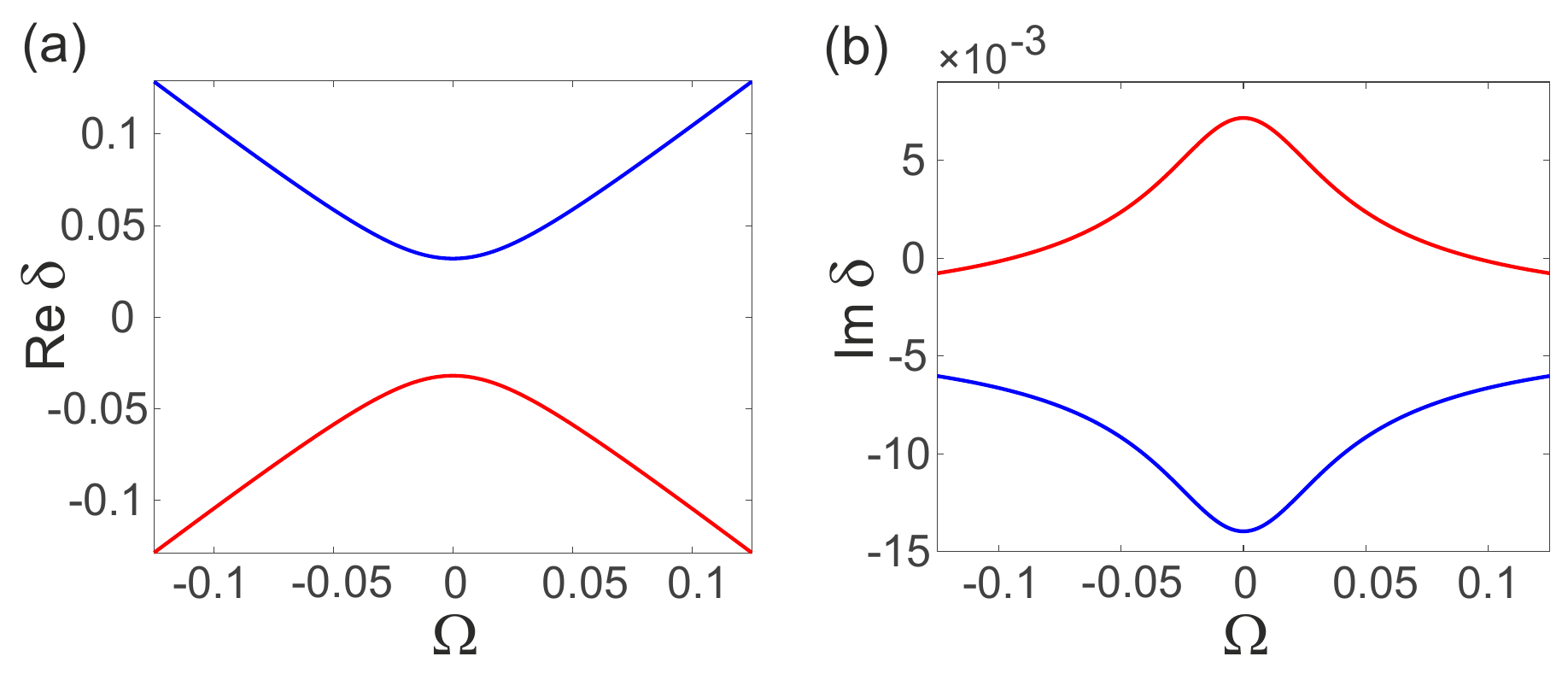}
\end{center}
\caption{ \label{fig8_new}
(Color online)
The dependencies of real (a) and imaginary (b) parts of the eigenfrequencies of the modes calculated in the framework of coupled mode theory are shown as the functions of the potential rotation velocity $\Omega$. The negative values of the imaginary part correspond to a growing mode.}
\end{figure}

The eigenfrequencies $\delta_1$ of the modes can be easily found analytically and read:
\begin{eqnarray}
\delta _1 = i \gamma_1 \pm \sqrt{4 \Omega^2+(1-i\alpha)^2\eta_{\pm 1}^{2} }. \label{mode_approx_LGD_CMA2} 
\end{eqnarray}
The eigenmodes have different linear growth rates due to the rotating spatially distributed linear gain. The dependencies of real and imaginary parts of the eigenfrequencies $\delta_1$ on the rotation velocity $\Omega$ are depicted in~Fig.~\ref{fig8_new}. In this particular case, we observe that there is now swap between the modes, and the same mode remains the fastest growing for all values of the angular velocity of the potential $\Omega$. Therefore, when $\Omega$ crosses zero, the change in the direction of polariton rotation occurs due to the corresponding change in the structure of the fastest growing mode. As a result, this change of the polariton velocity exhibits a smooth dependence  on the potential rotation velocity $\Omega$. It is worth mentioning that, indeed, the normalized angular momentum of the stationary polariton state matches well to the normalized angular momentum of the fastest growing linear mode in the whole range of $\Omega$,  see Fig.~\ref{fig6_new}(c).

Thus, we can conclude that close to the lasing threshold, the formation of the stationary state is determined by the fastest growing mode, suppressing the other modes. Let us remark that when the initial conditions are taken as high intensity noise, the resulting state becomes random, and we can observe, with different probabilities, the formation of the polariton states corresponding to different angular momenta.

Furthermore, it is important to note that at higher pump intensities, the impact of nonlinear effects on the structure of the growing field become significant. The normalized angular momenta obtained from the numerical simulations for for a stronger pump is shown in Fig.~\ref{fig6_new}(c) by the red squares. It is evident that they deviate significantly from the dependency predicted by the linear mode analysis. It is also instructive to compare the density distributions of the stationary states formed at different pumps, see Fig.~\ref{fig7_new}(d) and (e). It is seen that for relatively slow rotation velocities, a stationary state  resembles a standing wave with deep minima. This standing wave is formed by counter-propagating waves with similar amplitudes and slightly different frequencies, causing the standing wave to rotate with the potential. A finite angular momentum of a polariton state arises due to the difference of the amplitudes of the counter-propagating waves. At higher pumps, one of the waves becomes dominant, leading to an increase in the angular momentum of the state, see Fig.~\ref{fig6_new}(c). 

\section{Conclusion}

In conclusion, we developed a theory of the symmetry breaking in a polariton condensate formed in a stationary circular trap complemented with a weak rotating potential that helps hybridizing the modes of the condensate with different quantized angular momenta. It is demonstrated that in the vicinity of the lasing threshold, a ``winner takes it all'' scenario is realized, and the stationary state inherits the structure and normalized angular momentum of the fastest growing linear mode. We have also demonstrated that increase of the intensity of random initial noise can lead to the formation of stationary states different from the fastest growing one, with probabilities depending on the growth rates of these modes. 

We have identified the critical speeds of rotation of the potential that induce the change of the direction of the condensate rotation, and identified two corresponding switching mechanisms. In the first case, the structure of the fastest growing mode changes in such a way that at some potential rotation velocity the angular momentum of the mode changes its sign. In the second case, the fastest and the second fastest growing modes swap at threshold velocity.


\begin{acknowledgments}
A.V.Y. and I.A.S. acknowledge financial support from Icelandic Research Fund (Rannis, the project ``Hybrid polaritonics''), ``Priority 2030 Academic Leadership Program'' and ``Goszadanie no. 2019-1246''.
Work of E.S.S. was carried out within the state assignment in the field of scientific activity of the RF Ministry of Science and Higher Education (theme FZUN-2020-0013, state assignment of VlSU). 
A.V.K. and E.S.S. acknowledge Saint-Petersburg State University for
the financial support (research grant No.~94030557).
\end{acknowledgments}

\appendix

\setcounter{equation}{0}
\renewcommand*{\theequation}{A\arabic{equation}}

\section*{Appendix: Development of the perturbation theory for 
polaritons in a rotating trap}

Let us write the equation (\ref{master_scalar}) in the form
\begin{eqnarray}
\partial_t \psi = i \hat L \psi +i \hat P \psi + i G(\psi)
\label{app1}
\end{eqnarray}
where $\hat L = -\frac{i}{2} \nabla^2 \psi  +i(V +\frac{i}{2})\psi  - (i+\alpha) W_0$ accounts for the linear properties of the systems in the absence of the rotating potential $W_r$. The terms associated with the rotating potential  and the nonlinearity are $\hat P= -(i+\alpha )W_r$ and $G = (i+\alpha )(W_0+W_r) \frac{|\psi|^2}{1+|\psi|^2}+ i (h+i \tilde h)|\psi|^2\psi$, respectively.

To proceed, we represent the field as the series over the eigenmodes of the unperturbed problem 
$\partial_t \psi = i \hat L \psi  $. Because of the axial symmetry of the operator $\hat L$, the eigenmodes can be represented as $\psi_{m, q} = R_{m, q}(r)e^{i m \theta}e^{i\omega_{m, q} t}$, where $\omega_{m, q}$ represent the corresponding eigenfrequencies of the operator~$\hat L$: $\hat L \psi_{m, q}=\omega_{m, q}\psi_{m, q}$. The radial structure of the modes is described by the function $R_{m, q}$. The indices $q$ and $m$ are the radial and angular indices of the mode. In the further consideration we use the fact that $R_{m, q}=R_{-m, q}$ for the operator $\hat L$. For sake of convenience, we normalize the eigenfunctions $\int |\psi_{m, q}|^2 dx dy=1$.
Let us acknowledge here that the eigenvalues $\omega_{m, q}$ are complex, accounting also for the linear losses (gain) experienced by the modes. 

We look for a solution in the form 
\eqref{app_anzats} with coefficients $C_{m, q}(t)$ accounting for the temporal dynamics of the field. These coefficients evolve as $C_{m, q}=d_{m,q}e^{i\omega_{m, q} t }$, where $d_{m, q}$ are constants for the unperturbed problem. In the presence of perturbations, $d_{m, q}$ exhibits slow temporal dynamics. The validity of this approach relies on the condition that the perturbations are sufficiently weak, such that the characteristic evolution time $T_{ch}$ of the mode amplitudes $d_{m, q}$ satisfies the condition ${ \min( |\omega_{m, q}-\omega_{m', q'} |T_{ch}) \gg 1}$.

We further impose the condition $|\psi| \ll 1$, which corresponds to low density of polaritons when they weakly deplete the reservoir of incoherent excitons. This condition is required to reduce the nonlinearity associated with the depletion of the pump to a cubic nonlinearity $(i+\alpha )(W_0+W_r) |\psi|^2$. Let us also remark that under the aforementioned assumptions, the depletion of the rotating potential can be neglected, allowing us to express the nonlinear term in the following form:
\begin{eqnarray}
 G(\psi)=U |\psi|^2\psi, 
\label{app_nonl}
\end{eqnarray}
where $U(r)= i(h-W_0) -(\tilde h +\alpha W_0)$ is the complex pseudo-potential.

To develop the perturbation theory, we start with the linear inter-mode interactions induced by the rotating potential. By substituting (\ref{app_anzats}) into (\ref{app1}) and projecting onto the $\psi_{m, q}$ modes, we derive the equation for $d_{m, q}$ in the form
\begin{multline}
 \dot C_{m, q}= i\omega_{m, q} C_{m, q} \\ +\sum_{m', q'} C_{m', q'} \int \psi_{m', q'} \psi_{m, q}^{*} W_r dx dy .
\label{app2}
\end{multline}
We consider the rotating potential of the form $W_r=(i+\alpha)f(r) \cos [\Delta l (\theta-\Omega t) ]$, where $f(r)$ describes the radial dependency of the potential. Let us remind that $\Delta l$ is the number of maxima of the potential along the angular coordinate, and $\Omega$ is the potential rotation  velocity. We make the derivation for the case of $\Delta l=1$, which can be easily generalized to arbitrary $\Delta l$. Subsequently, the integrals can be calculated as
\begin{widetext}
$$ \int \psi_{m', q'} \psi_{m, q}^{*} W_r dx dy=\frac{1}{2} \int r R_{m, q}^{*} R_{m', q'} f(r) dr \int e^{i(m'-m+1)\theta-i\Omega t
}+e^{i(m'-m-1)\theta+i\Omega t} d\theta.$$ 
\end{widetext}
This term is nonzero only for $m'=m-1$ and $m'=m+1$. 
Then the equations for $C_{m, q}$ can be reduced to 
\begin{widetext}
\begin{eqnarray}
 \dot C_{m, q}= i\omega_{m, q} C_{m, q}+(i+\alpha)\sum_{q'}  \sigma_{m, +, q, q'} e^{i\Omega t} C_{m+1, q'}+\sigma_{m, -, q, q'} e^{-i\Omega t} C_{m-1, q'}, 
\label{app3}
\end{eqnarray}
\end{widetext}
where $\sigma_{m, \pm, q, q'}=\pi \int r R_{m, q}^{*} R_{m \pm 1, q'} f(r) dr$. 

Typically, only a few modes exhibit negative or low positive losses, which aligns well with the results presented in the main part of the paper. To obtain the coefficients required for the perturbation theory, we conducted several numerical experiments. Initially, let us determined the eigenfrequencies of the polariton modes, specifically for the case of small $\alpha$ where the gain does not significantly impact the mode structure. Under this condition, the real parts of the eigenfrequencies remain independent of~$\alpha$. Therefore, we choose $\alpha=0.26$, slightly below the smallest lasing threshold, and perform the simulation starting with the initial conditions in the form of $\psi=a\sum_{m=0}^{4} e^{-r^2/r_0^2}e^{i m \theta}$, where $a=0.001$ is the small amplitude and $r_0=1$ is the radius of the initial field distribution. The rotating potential is set to zero. 

\begin{figure}[tb!]
\begin{center}
\includegraphics[width=\linewidth]{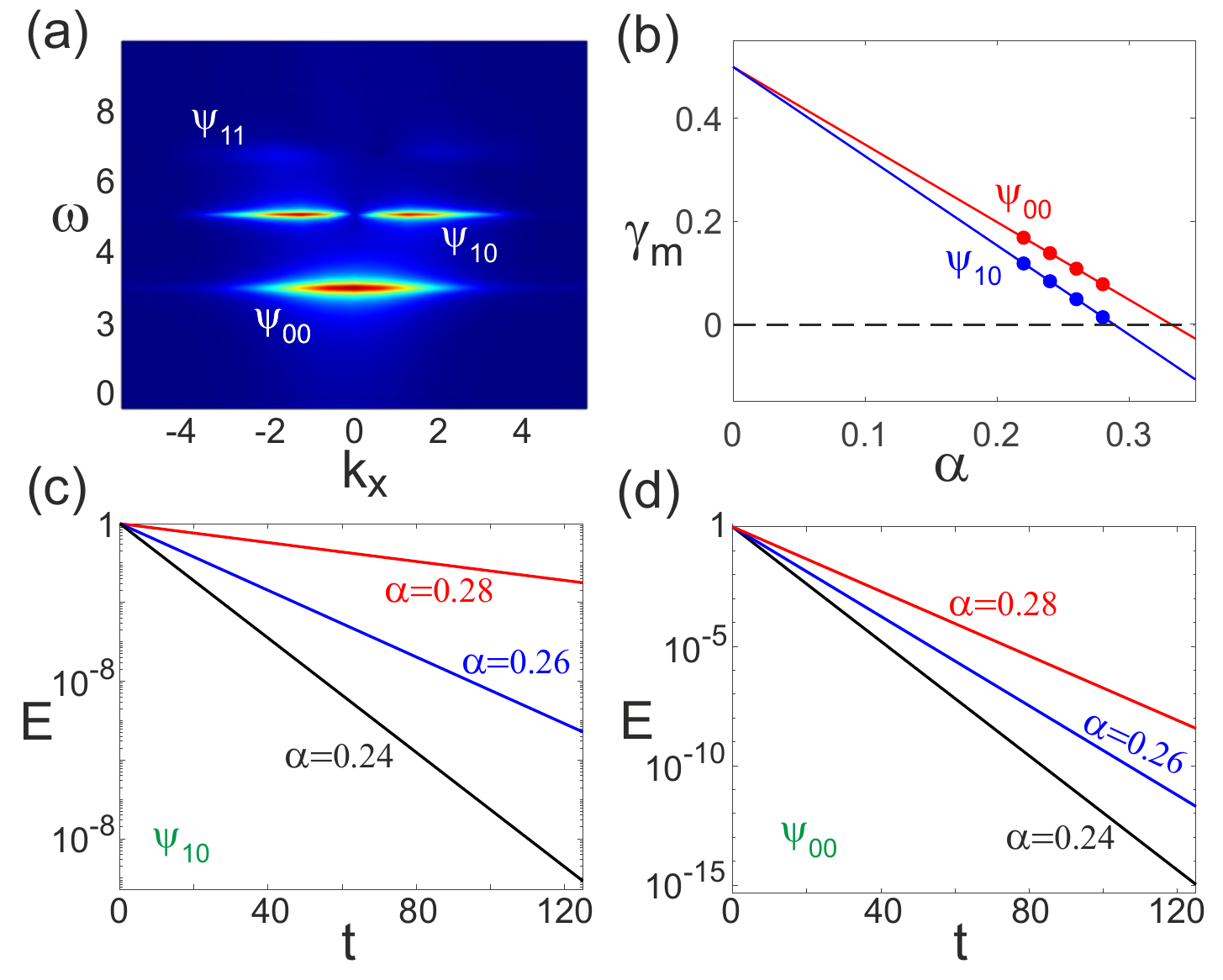}
\end{center}
\caption{ \label{figApp1}
(Color online)
Cross section of the Fourier representation $\psi(k_x, k_y=0, \omega)$ of the field $\psi(x, y, t)$ is shown in panel (a). The patterns identified as the modes with $m=0$,~$q=0$, $m=1$,~$q=0$ and $m=1$,~$q=1$ are marked as $\psi_{00}$, $\psi_{10}$ and $\psi_{11}$, respectively. The calculated dependencies of the decay rates of the modes $\psi_{00}$ and $\psi_{10}$ as functions of $\alpha$ are shown in panel (b). The  solid circles are the calculated points, the thin lines are the guide for eye. The dashed line marks the zero of the decay rates. The temporal dependencies of the energies of the modes $\psi_{01}$ and $\psi_{00}$ on time are shown in panels (c) and (d) for different values of $\alpha$. The time axis is chosen such that at $t=0$ the only one mode from the initial field distributions survives. For sake of convenience, the energies are scaled such that at $t=0$ they all are equal to unity. Please see text for more details.}
\end{figure}

The results of the numerical simulations are presented in Fig.~\ref{figApp1}(a) displaying the Fourier representation $\psi(k_x, k_y, \omega)$ of the field $\psi(x, y, t)$. In this panel, one can easily identify the pattern corresponding to the mode with $m=0$, $q=0$ marked as $\psi_{00}$. At the higher frequency, there is a pattern corresponding to the mode $\psi_{10}$ with $m=1$,~$q=0$. By calculating the positions of the spectral lines, we determine the eigenfrequencies of the modes.

It is worth noting that for the chosen parameters the width of the spectral line of the mode $\psi_{10}$ is smaller compared to that of the mode $\psi_{00}$, indicating a higher quality factor for the modes~$\psi_{\pm 1 0}$. Additionally, a faint pattern corresponding to the mode $\psi_{11}$ is observed at an even higher frequency, but its quality factor is significantly lower. Let us remark that the eigenfrequency of this mode is close to the maximum height of the confining potential, suggesting that radiative losses are expected to be very high for this mode and for the modes with higher frequencies. 

By conducting numerical simulations and monitoring the decay of the polariton number
$E=\int |\psi|^2 dx dy$ over time, it is straightforward to determine the mode decrements as functions of the parameter $\alpha$. By choosing appropriate symmetry of the initial conditions, it is possible to distinguish the modes with different $m$. Thus, the initial distribution contain the modes with fixed $m$ but different $q$. However the modes with different $q$ have different decay rate, and eventually, only the mode with the slowest decay rate survives. This way, we estimate the mode decrements as functions of $\alpha$ for the modes $\psi_{10}$ and $\psi_{00}$, see Fig.~\ref{figApp1}(b). To demonstrate that we conduct the estimation at times when 
the field contains only one mode, we plotted the dependencies of the polariton number $E$ on time, see Fig.~\ref{figApp1}(c,d). One can see that these dependencies exhibit linear behavior in the logarithm scale.  

As expected for a weakly dissipative system, the decay rates of the modes depend linearly on the parameter $\alpha$, which represents the ratio of the imaginary part of the potential to its real part. This it is possible to extrapolate the dependencies for finding the effective losses for the parameter $\alpha=0.33$ used in the numerical simulations. It should be noted that the mode decrements become negative at some threshold $\alpha$, indicating the mode growth. This method also provides the ability  to identify the mode with longest lifetime among the modes with a fixed angular index~$m$. 

Finding the modes with the second smallest decay rate is more challenging. However, an estimation of their decay rate can be obtained by examining the beating between the modes. The decay rate of this beating provides the difference between the decay rates of the modes with the first and the second smallest decay rates. By conducting this estimation, we discovered that the decay rate of the mode $\psi_{11}$ is more than an order of magnitude higher that the decay rate of the mode $\psi_{10}$. This observation supports the assumption that, in the first approximation, the dynamics of the system can be described in terms of the amplitudes of the modes $\psi_{00}$ and $\psi_{\pm1 0}$.   

Thus, we disregard all modes except the three modes with $m=\pm 1$, $m=0$ and $q=0$. Subsequently, the equations for the coefficients $d_{\pm1}$, $d_0$ take the following form:
\begin{subequations}
\begin{eqnarray}
&& \dot C_{-1}= i\omega_{1} C_{-1}+ \eta e^{i\Omega t} C_{0} , \label{app4_1} \\
&&\dot C_{0}= i\omega_{0} C_{0}+ \eta \left( e^{-i\Omega t} C_{-1} +e^{i\Omega t} C_{1}\right) , \label{app4_2}\\
&&\dot C_{1}= i\omega_{1} C_{1} +\eta e^{-i\Omega t} C_{0}. 
\label{app4_3}
\end{eqnarray}
\end{subequations}
Here and everywhere below we omitted the second index $q$ for it is always zero. We also used the equalities ${\omega_{m}=\omega_{-m}}$ and ${R_{m}=R_{-m}}$ that follow from degeneracy of the modes with angular indices $\pm m$. This is why $\sigma_{-1, +}=\sigma_{0, -}=\sigma_{0, +}=\sigma_{1, -}$ and we denote this as $\eta_{10}$.

In order to use the coupled mode approach, it is necessary determinate the values of the scattering rates $\eta_{10}$. These values can be found from 2D numerical simulations. We choose $\alpha$ slightly below the lasing threshold for $\psi_{10}$ mode and take the initial distribution in the form $\psi=a e^{-r^2/w^2}e^{i\theta}$. Then conduct numerical simulations in the presence of the conservative rotating potential and calculate the temporal spectra of the field. 

\begin{figure}[tb!]
\begin{center}
\includegraphics[width=\linewidth]{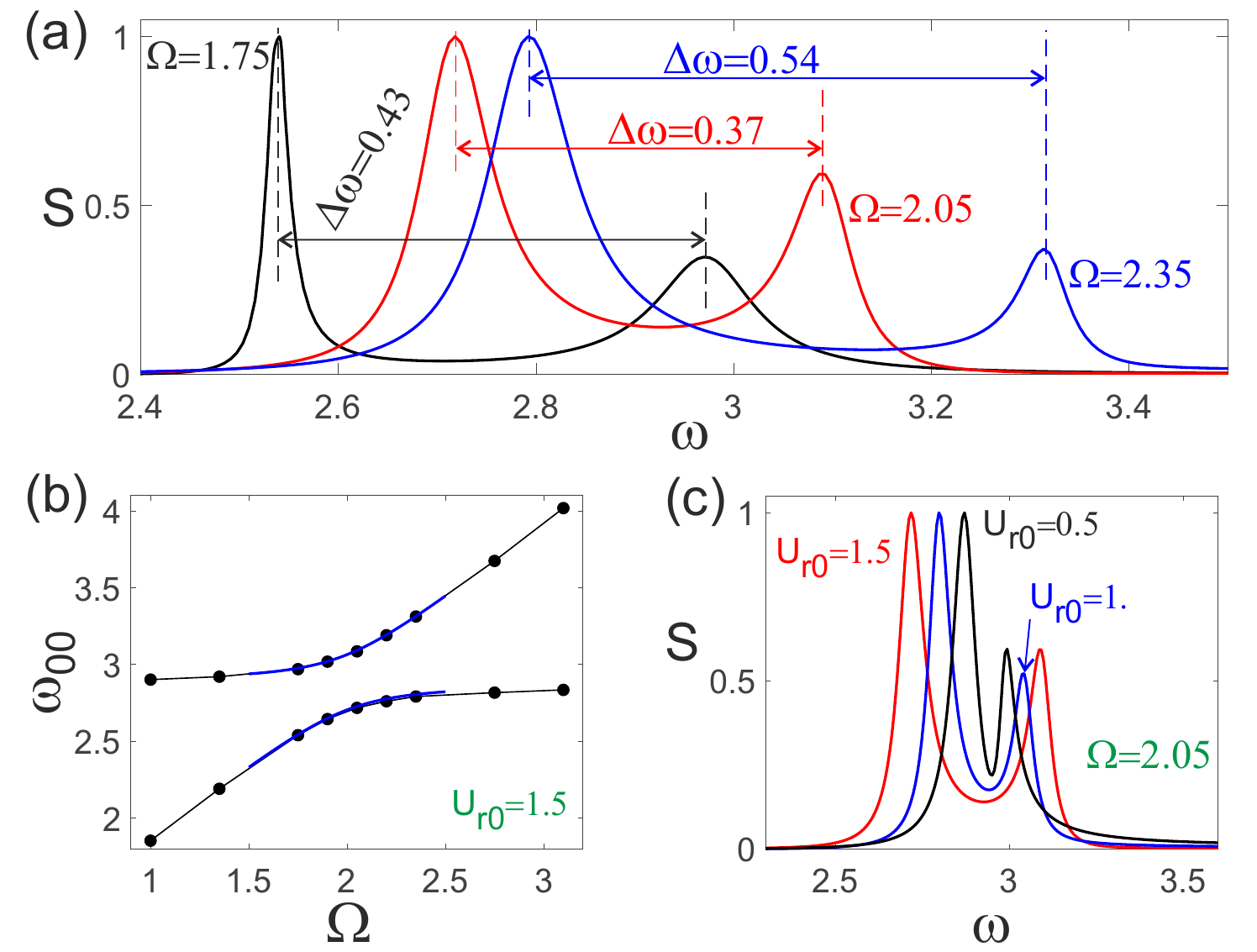}
\end{center}
\caption{ \label{figApp2}
(Color online)
The temporal spectra of the field are shown in panel (a) for different potential rotation velocities $\Omega$. The frequency differences $\Delta \omega$ between the peaks are also given in the panel. Panel (b) shows the positions of the spectral peaks in the vicinity of the eigenfrequency of $\psi_{00}$ state as a functions of the angular velocity of the rotating potential $\Omega$. The black solid circles are the data extracted from 2D simulations, the blue curves are the fit by formula~\eqref{disp1_1}. The think black lines are guide for the eye. Panel (c) show the spectra calculated for the different amplitudes of the potential rotating at the velocity $\Omega=2$.}
\end{figure}

The presence of the rotating potential with $\Delta l=1$ causes the coupling of the mode with $m=1$ to the mode with $m=0$. This process is most effective when the rotation velocity $\Omega$ matches the frequency difference between the modes $\psi_{00}$ and $\psi_{10}$. Indeed, when the frequency is close to $2$, we observe the emergence of two spectral peaks with the frequencies close to the eigenfrequency of the $\psi_{00}$ mode, see Fig.~\ref{figApp2}(a). 

These peaks correspond to two hybrid modes formed by the mixing of $\psi_{10}$ and $\psi_{00}$ states. It is worth noting that each of the hybrid modes has two spectral lines, one near the frequency of the pure $\psi_{10}$ state and the other near the frequency of the $\psi_{00}$ state. Because of the technical reasons, it is easier to estimate the spectral lines at the frequencies close to the eigenfrequency of $\psi_{00}$ state.

The positions and the separation between the spectral lines of the hybrid modes are dependent on the rotation velocity of the potential, see Fig.~\ref{figApp2}(a) and (b). To confirm that the two peaks are indeed associated with the splitting caused by the rotating potential, we measure the spectra for different amplitudes of the rotating potential. The results depicted in Fig.~\ref{figApp2}(c) show that, indeed, the frequency difference between the peaks decreases with the decrease of the rotating potential amplitude. 

In the vicinity of the resonance frequency, this dependency can be approximated by formula (\ref{disp1_1}). To estimate the scattering rate, we can disregard the losses in (\ref{disp1_1}) and fit the numerical data by the analytical expression (\ref{disp1_1}) choosing the appropriate $\eta$. The fit, shown by the thicker blue line in Fig.~\ref{figApp2}(b), demonstrates good agreement between the analytical and numerical results.

Let us now turn to the perturbation induced by the nonlinear term. Looking for the solution in the form
$$\psi=C_{-1} R_{1}e^{-i\theta}+C_{0} R_{0}+C_{1} R_{1}e^{i\theta} ,$$
we can calculate the nonlinear term and project it onto the eigenfunctions $\psi_{\pm 1}$ and $\psi_0$. The projections $Y_{\pm 1}$, $Y_{0}$ on the modes $m=-1$, $m=0$ and $m=1$ are given by the following expressions
\begin{widetext}
\begin{subequations}
\begin{eqnarray}
&&Y_{-1}= \left( \epsilon_{11}|C_{-1}|^2+2\epsilon_{11}|C_{1}|^2 + \epsilon_{01} |C_0|^2\right)C_{-1}+\tilde \epsilon_{01}C_{0}^2C_{1}^{*} ,\label{app5_1} \\
&&Y_{0}=\left( \epsilon_0 |C_0|^2+ 2\epsilon_{01}|C_{-1}|^2+ \epsilon_{01}|C_{1}|^2 \right)C_0 + \tilde \epsilon_{01}^{*} (C_0+C_0^{*})C_{-1}C_{1} , \label{app5_2}\\
&&Y_{1}= \left( \epsilon_{11}|C_{1}|^2+2\epsilon_{11}|C_{-1}|^2 + \epsilon_{01} |C_0|^2\right)C_{1}+\tilde \epsilon_{01}C_{0}^2C_{-1}^{*}, \label{app5_3}
\end{eqnarray}
\end{subequations}
\end{widetext}
where the nonlinear interaction constants are $\epsilon_{11}=2\pi\int U r |R_1|^4 dr$, $\epsilon_{01}=4\pi\int U r |R_0|^2| R_1|^2 dr$, $\epsilon_{0}=2\pi\int U r |R_0|^4 dr$ and $\tilde \epsilon_{01}=2\pi\int U r R_0^2 R_1^{*2} dr$.

It is worth noting that the first term in (\ref{app5_1}) has the frequency close to the frequency of the mode $\omega_1$ provided that, according to derivation, the mode amplitudes $C$ can be represented as $C_m=d_m e^{i\omega_m t}$, where $d_m$ is a slowly varying function of time. However, the frequency of the second term can be estimated as $2\omega_0-\omega_1$. This would cause the dynamics of the mode amplitude $d_{-1}$ at the frequency $2|\omega_0-\omega_1|$ which is, obviously, not small compared to the detuning of the modes eigenfrequencies. Therefore, this term cannot be included in the equation for $C_{1}$ but, if necessary, can be taken into account through an additional small correction to the field. Similarly, it can be observed that the last terms in the expressions $Y_0$ and $Y_{1}$ do not contribute to the equations for the amplitudes $C_0$ and $C_1$. Thus, accounting for the nonlinear interaction, the equations for the mode amplitudes can be written as follows:
\begin{widetext}
\begin{subequations}
\begin{eqnarray}
&&\dot C_{-1}= i\omega_{1} C_{-1}+ (i+\alpha)\eta_{10} e^{i\Omega t} C_{0} + \left( \epsilon_{11}|C_{-1}|^2+2\epsilon_{11}|C_{1}|^2 + \epsilon_{01} |C_0|^2\right)C_{-1} \label{app6_1}, \\
&&\dot C_{0}= i\omega_{0} C_{0}+ (i+\alpha)\eta_{10} \left( e^{-i\Omega t} C_{-1} +e^{i\Omega t} C_{1}\right) +\left( \epsilon_0 |C_0|^2+ \epsilon_{01}|C_{-1}|^2+ \epsilon_{01}|C_{1}|^2 \right)C_0  \label{app6_2}, \\
&&\dot C_{1}= i\omega_{1} C_{1} +(i+\alpha)\eta_{10} e^{-i\Omega t} C_{0}+\left( \epsilon_{11}|C_{1}|^2+2\epsilon_{11}|C_{-1}|^2 + \epsilon_{01} |C_0|^2\right)C_{1}. 
\label{app6_3}
\end{eqnarray}
\end{subequations}
\end{widetext}

Now changing the variables $A_1=C_1 \exp^{-i(\omega_{0r} +\Omega)t}$, $A_0=C_0 \exp^{-i\omega_{0r} t}$ and $A_{-1}=C_{-1} \exp^{-i(\omega_{0r} -\Omega)t}$, where $\omega_{0r}= \text{Re}\,\omega_0$, we arrive to the equations for the amplitudes $A$:
\begin{widetext}
\begin{subequations}
\begin{eqnarray}
&&\dot A_{-1}= (-\gamma_1 + i\Delta_1+i\Omega) A_{-1}+ (i+\alpha) \eta_{10} A_{0} + \left( \epsilon_{11}|A_{-1}|^2+2\epsilon_{11}|A_{1}|^2 + \epsilon_{01} |A_0|^2\right)A_{-1} , \label{app7_1} \\
&&\dot A_{0}=-\gamma_0 A_{0} + (i+\alpha)\eta_{10}\left(  A_{1} + A_{-1} \right) +\left( \epsilon_0 |A_0|^2+ \epsilon_{01}|A_{-1}|^2+ \epsilon_{01}|A_{1}|^2 \right)A_0 , \label{app7_2} \\
&&\dot A_{1}=(-\gamma_1 +i\Delta_1 - i\Omega) A_{1} + (i+\alpha)\eta_{10} A_{0}+\left( \epsilon_{11}|A_{1}|^2+2\epsilon_{11}|A_{-1}|^2 + \epsilon_{01} |A_0|^2\right)A_{1}., 
\label{app7_3}
\end{eqnarray}
\end{subequations}
\end{widetext}
where $\gamma_1= \text{Im} \, \omega_1$, $\gamma_0= \text{Im} \, \omega_0$ and $\Delta_1= \text{Re} \, \omega_1  - \omega_{0r}$.

\begin{figure}[tb!]
\begin{center}
\includegraphics[width=\linewidth]{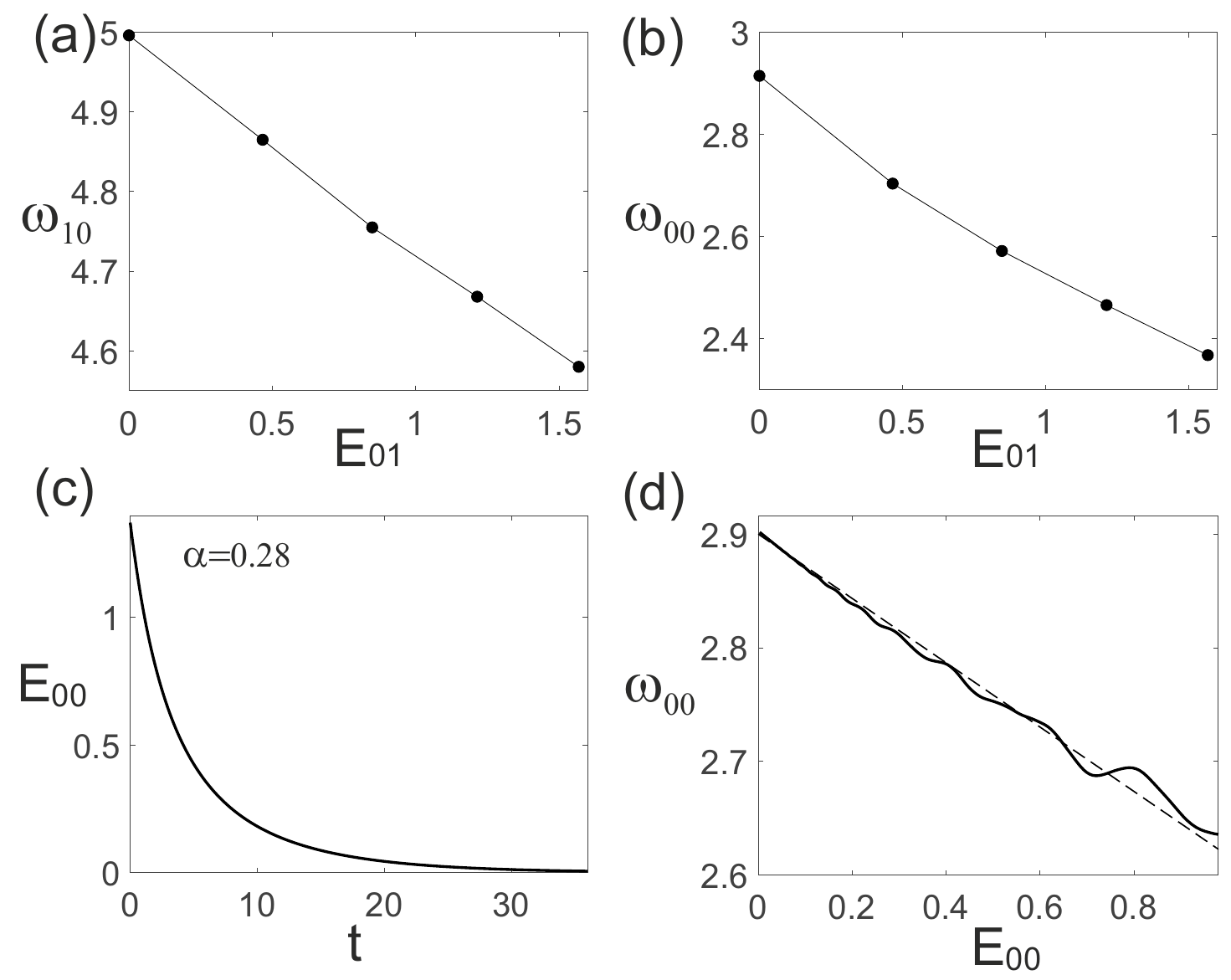}
\end{center}
\caption{ \label{figApp3}
(Color online)
The dependency of the frequency $\omega_{10}$ of the  stationary state $\psi_{10}$ on its polariton number 
$E_{10}$ is shown in panel (a). The frequency of the weak perturbation having the structure of the state $\psi_{00}$ nestling on the stationary state $\psi_{10}$ is shown in (b) as a function of the total number of the polaritons $E_{10}$ in the sate $\psi_{10}$. The temporal evolution of the polariton number  
$E_{00}$ of the state $\psi_{00}$ is shown in panel (c) for the parameter $\alpha$ below the lasing threshold. The instantaneous frequency $\omega_{00}=\frac{\partial_t Q}{Q}$ of the decaying state $\psi_{00}$ is shown in (d) as a function of the total number of polaritons $E_{00}$ in the state $\psi_{00}$. The parameter $Q$ is defined as $Q=\int \psi dx dy$. The thinner dashed line in this panel is the linear fit for the numerically calculated curve.  
}
\end{figure}

Next, we need to determining the values of the nonlinear interaction constants. We start with estimating the imaginary parts of $\epsilon$'s. For this purpose, we perform numerical simulations of the problem without the rotating potential, while maintainig $\alpha$ above the lasing threshold for for the mode $\psi_{10}$ but below the lasing threshold of the mode $\psi_{00}$. As a result, only the state $\psi_{10}$ can form in the system. The frequency of the stationary state can be found from the position of the maximum of its temporal spectrum. The 
polariton number of the stationary state $E_{10}$ can be controlled by  changing the value of $\alpha$. This way it is possible to find the dependency of the frequency of the stationary state on its polariton number. This dependency is shown in Fig.~\ref{figApp3}(a). It is seen that the dependency is quite linear which proofs that the nonlinearity can be accurately approximated by a cubic nonlinearity. The slope of the line gives the value of the coefficient $\text{Im} \, \epsilon_{11} $.

To find the coefficient $\text{Im} \, \epsilon_{01}$ we perturb the stationary state $\psi_{11}$ by the state $\psi_{00}$ of very low intensity. Then we measure the frequency of the state $\psi_{00}$ which depends on the polariton number of the state $\psi_{10}$. The frequencies of the states $\psi_{00}$ and $\psi_{10}$ are well resolved and so the frequency of the state $\psi_{00}$ can be measured easily. The dependency of the frequency $\omega_{00}$ of the small perturbation $\psi_{00}$ on the polariton number $E_{10}$ of the stationary nonlinear state $\psi_{10}$ is shown in Fig.~\ref{figApp3}(b). As it is expected the dependency is also close to one in the linear case. 

To determine the coefficient $\epsilon_{00}$ characterizing the nonlinear self-action of the mode $\psi_{00}$, we take the initial conditions in the form of the field with the angular index $m=0$ with rather high intensity. To avoid the excitation of the state $\psi_{10}$, we set $\alpha$ below the lasing threshold. In these circumstances, the polariton number of the state $\psi_{00}$ decays quasi-exponentially, see Fig.~\ref{figApp3}(c). The change in intensity of the state leads to variations in its frequency and decrement. The instantaneous frequency can be obtained by calculating the function $\frac{\partial Q}{Q}$, where $Q=\int \psi dx dy$. The dependency of the frequency $\omega_{00}$ on the polariton number $E_{00}$ of the mode $\psi_{00}$ is shown in Fig.~\ref{figApp3}(d). The dependency is also close to one in the linear case. The deviation can be explained by the contribution of the modes with higher radial indices. The slope of the dependency allows us to determine~$\text{Im} \, \epsilon_{00}$.

Let us note that by measuring the decays rate of the modes, one can obtain  the values of the real part of $\epsilon$, which quantify the strength of the nonlinear losses. For the chosen parameters, the real parts of $\epsilon$ can be approximated by the formula ${\text{Re} \, \epsilon = \alpha \text{Im} \, \epsilon}$, since the dominant nonlinearity arises from pump depletion. Furthermore, it is worth mentioning that the same approach described above can be used to determine the parameters for the coupled mode approximation in the case of a potential with~$\Delta l = 2 $.

\bibliography{main_text_yulin}

\begin{thebibliography}{40}
\bibitem [{Rea()}]{RealParamValues}%
  \BibitemOpen
  \href@noop {} {}\bibinfo {note} {The following values of dimensional
  parameters in Eq.~(2) correspond to the values of dimensionless parameters
  used for simulations. The effective polariton mass is $m^* = 5 \times 10^{-5}
  m_{e}$, where $m_e$ is the free electron mass in vacuum, the polariton and
  reservoir excitons decay rates are $\Gamma _p = 0.06 \, \text{ps} ^{-1}$ and
  $\Gamma _r = 0.015 \, \text{ps}^{-1}$, respectively, the stimulated
  scattering rate is~$ g_1 = 0.006 \, \text{ps}^{-1} \, \mu \text{m}^2$. The
  polariton-polariton and polariton-exciton scattering constants are estimated
  as $H \approx 1.18 \, \mu \text{eV} \, \mu {\text{m}^2} $ and $\hbar g_2
  \approx 12 \, \mu \text{eV} \, \mu {\text{m}^2}$, respectively, the nonlinear
  losses are $\tilde{H} \approx 0.1 \, \text{ns}^{-1} \,
  \mu\text{m}^2$.}\BibitemShut {Stop}%
\end{thebibliography}
 
\end{document}